\documentclass{article}

\usepackage{arxiv}

\usepackage[utf8]{inputenc} 
\usepackage[T1]{fontenc}    
\usepackage[colorlinks=true, citecolor=blue, linkcolor=blue]{hyperref}       
\usepackage{url}            
\usepackage{booktabs}       
\usepackage{amsfonts}       
\usepackage{nicefrac}       
\usepackage{microtype}      
\usepackage{lipsum}
\usepackage{graphicx} 
\usepackage{natbib}
\usepackage{paralist}

\title{Martian Year 34 Column Dust Climatology from Mars Climate Sounder Observations: Reconstructed Maps and Model Simulations}

\title{Martian Year 34 Column Dust Climatology from Mars Climate Sounder Observations: Reconstructed Maps and Model Simulations}

\newcommand{\lmd}{%
Laboratoire de M\'et\'eorologie Dynamique (LMD/IPSL), Sorbonne Université, \\
Centre National de la Recherche Scientifique, {\'E}cole Polytechnique, \\
{\'E}cole Normale Supérieure, Paris, France
}

\author{
  Luca Montabone\\
  Space Science Institute, Boulder, CO, USA\\ and\\ \lmd\\
   \And
  Aymeric Spiga \\
  \lmd\\ and\\ Institut Universitaire de France, Paris, France\\
   \And
 David M. Kass and Armin Kleinb\"ohl \\
  Jet Propulsion Laboratory, Caltech Institute of Technology, Pasadena, CA, USA\\
   \AND
  Fran\c cois Forget and Ehouarn Millour \\
  \lmd\\
  \\
  Corresponding author: Luca Montabone (\texttt{lmontabone@spacescience.org})
}

\begin{document}
\maketitle

%
%

%
%


\begin{abstract}
We have reconstructed longitude-latitude maps of column dust optical depth (CDOD) for Martian year (MY) 34 (May 5, 2017 --- March 23, 2019) using observations by the Mars Climate Sounder (MCS) aboard NASA's Mars Reconnaissance Orbiter spacecraft. Our methodology works by gridding standard and newly available estimates of CDOD from MCS limb observations, using the ``iterative weighted binning'' methodology. In this work, we reconstruct four gridded CDOD maps per sol, at different Mars Universal Times. Together with the seasonal and day-to-day variability, the use of several maps per sol allows to explore also the daily variability of CDOD in the MCS dataset, which is shown to be particularly strong during the MY 34 equinoctial Global Dust Event (GDE). Regular maps of CDOD are then produced by daily averaging and spatially interpolating the irregularly gridded maps using a standard ``kriging'' interpolator, and can be used as ``dust scenario'' for numerical model simulations. In order to understand whether the daily variability of CDOD has a physical explanation, we have carried out numerical simulations with the ``Laboratoire de M\'et\'eorologie Dynamique'' Mars Global Climate Model. Using a ``free dust'' run initiated at $L_s \sim 210^\circ$ with the corresponding kriged map, but subsequently free of further CDOD forcing, we show that the model is able to account for some of the observed daily variability in CDOD. The model serves also to confirm that the use of the MY 34 daily-averaged dust scenario in a GCM produces results consistent with those obtained for the MY 25 GDE.
\end{abstract}

\section*{Plain Language Summary}

Large dust storms on Mars have dramatic impacts on the entire atmosphere, but may also have critical consequences for robotic and future human missions. Therefore, there is compelling need to produce an accurate reconstruction of their spatial and temporal evolution for a variety of applications, including to guide Mars climate model simulations. The recently ended Martian year 34 (May 5, 2017 – March 23, 2019) represents a very interesting case because an extreme dust event occurred near the time of the northern autumn equinox, consisting of multiple large dust storms engulfing all longitudes and most latitudes with dust for more than 150 Martian days (``sols’’). We have used satellite observations from the Mars Climate Sounder  instrument aboard NASA’s Mars Reconnaissance Orbiter to reconstruct longitude-latitude maps of the opacity of the atmospheric column due to the presence of dust at several times in each sol of Martian year 34. These maps allow us to analyze the seasonal, day-do-day, and day-night variability of dust in the atmospheric column, which is particularly intense during the extreme dust event. We have also used simulations with a Mars climate model to show that the strong day-night variability may be partly explained by the large-scale circulation.

\section{Introduction}

Martian dust aerosols are radiatively active, and the dust cycle --- lifting, transport, and deposition --- is considered to be the key process controlling the variability of the Martian atmospheric circulations on a wide range of time scales \citep[see e.g. the recent review by ][ and references therein]{Kahr:17}. Dust storms are the most remarkable manifestations of this cycle, and one of the most crucial weather phenomena in need to be studied to fully understand the Martian atmosphere. 

Martian dust storms are:
\begin{inparaenum}
\item a source of strong atmospheric radiative forcing, and alteration of surface energy budget \citep[e.g. ][]{Stre:19};
\item a major component of the atmospheric inter-annual, seasonal, day-to-day, and daily variability [e.g. Kleinb\"ohl et al., this issue]; 
\item a way to redistribute dust on the planet via long-range particle transport;
\item a key element to also understand longer-term surface changes, such as, for example, the aeolian erosion, and the retreat of the seasonal ice cap;
\item a strong disturbance of temperature and density propagating from the lower to the upper atmosphere, including the lower thermosphere, the ionosphere, and the magnetosphere [e.g. Girazian et al., and Xiaohua et al., this issue]; 
\item a cause of increase of loss of chemical species via escape \citep[][and Xiaohua et al., this issue]{Fedo:18,Heav:18}; 
\item a source of hazards for spacecraft Entry, Descending and Landing (EDL) manoeuvres, for operations by solar-powered surface assets, and for future robotic and human exploration \citep[e.g. ][]{Levi:18book}.
\end{inparaenum}
Dust storms on Mars can be studied by using a variety of approaches: analysis of observations from satellites and landers/rovers, numerical simulations from Global Climate Models (GCM), and data assimilation techniques.

One of the most dramatic and (so far) unpredictable event linked to Martian dust storms is the onset of a Global Dust Event --- hereinafter GDE. In the literature, these events are also named ``Planet-Encircling Dust Storms'' \citep[e.g.][]{Cant:07}, or ``Global Dust Storms''. Here we choose the terminology ``Global Dust Event'' \citep[already discussed and used in ][]{Mont:18book} because 
\begin{inparaenum} 
\item even large regional storms can inject dust high enough in the atmosphere, which eventually encircles the planet, and
\item these dust events are usually characterized by several storms occurring simultaneously, or one after the other one in rapid succession. 
\end{inparaenum}

In the last Martian decade (Martian Year --- hereinafter MY --- 25 to 34), spanning nearly two Earth decades from 2000 to 2019, three GDEs occurred: an equinoctial event in MY 25, a solstitial event in MY 28, and another equinoctial event in MY 34, starting only few mean solar days --- sols --- after the corresponding onset of the MY 25 event. GDEs inject a large amount of dust particles in the Martian atmosphere, strongly modify the thermal structure and the atmospheric dynamics during several months \citep[i.e. several tens degrees of solar longitude --- $L_s$ ---, see e.g.][]{Wils:96,Mont:05luca}, and impact the Martian water cycle and escape rate \citep{Fedo:18,Heav:18}. Similar events were previously observed in Martian Years 1, 9, 10, 12, 15, and 21 \citep{Mart:93leo,Cant:07,Mont:18book,Sanc:19storm}. The inter-annual variability of GDEs is irregular, and likely controlled at the first-order by the redistribution of dust on Mars over the timescale of a few years \citep{Mulh:13,Vinc:15}.

The latest equinoctial GDE had its initial explosive growth in early northern fall of MY 34 ($L_s$ approximately in the range $185-190^{\circ}$, i.e. late May 2018 -- early June 2018). A regional dust storm started near the location of the Mars Exploration Rover ``Opportunity''. The visible opacity quickly reached a very high value of~$10.8$, which led to the end-of-mission of the Opportunity rover, with last communication received on June 10, 2018. The regional dust storm then moved southward along the Acidalia storm-track, and expanded both in the northern hemisphere from eastern Tharsis to Elysium (including the location of the Mars Science Laboratory ``Curiosity'' rover and the landing site of InSight), and towards the southern hemisphere \citep[][]{Mali:18rds}. 

The evolution of this MY 34 GDE in summer 2018 has been closely monitored by three NASA's orbiters, including the Mars Reconnaissance Orbiter (MRO) and its Mars Climate Sounder (MCS) instrument [Kass et al. this issue], two ESA's orbiters (Mars Express and ExoMars Trace Gas Orbiter), the ISRO's Mangalyaan orbiter, and ground-based telescopes \citep{Sanc:19storm}. It has also been observed in details from the surface by the Curiosity rover, which could still operate in dust-storm conditions thanks to its nuclear-powered system. From meteorological observations carried out aboard Curiosity with the Rover Environmental Monitoring Station, \citet{Guze:18} concluded that the local optical depth reached 8.5, the incident total UV solar radiation at the surface decreased by 97\%, the diurnal range of air temperature decreased by 30 K, and the semidiurnal pressure tide amplitude increased to 40 Pa. Curiosity did not witness dust being lifted within the Gale Crater site, which indicates that the increase in dust loading at its location corresponds to dust particles transported in the Gale Crater area by large-scale and regional winds. 

Beyond the undoubtedly interesting GDE, MY 34 also features the development of an unusual late-winter large regional storm whose peak CDOD is only rivaled by that in MY 26, reaching 75\% of its peak value. It is, however, reminiscent of the two global events that were successively monitored by the Viking landers in 1977 at $L_s = 205^{\circ}$ and $L_s = 275^{\circ}$ \citep{Ryan:79,Zure:82}. Overall, therefore, MY 34 represents a unique year for studies linked to the onset/evolution of dust storms, and their impact on the entire Martian atmospheric system. Consequently, there is a compelling need to produce an accurate reconstruction of the spatial and temporal evolution of the dust optical depth in MY 34, particularly covering the GDE, but also putting the unprecedented weather measurements acquired by the InSight lander during the late-winter regional storm in the global context \citep{Spig:18insight}. 

\citet{Mont:15} have developed a methodology to grid values of column dust optical depth (CDOD) retrieved from multiple polar orbiting satellite observations, such as NASA's Mars Global Surveyor (MGS), Mars Odyssey (ODY), and MRO. Using this methodology (a combination of ``Iterative Weighted Binning'' --- IWB --- and kriging spatial interpolation), they were able to produce a multi-annual dataset of daily CDOD maps extending from MY 24 to MY 33, which is publicly available on the Mars Climate Database (MCD) project webpage at \url{http://www-mars.lmd.jussieu.fr/} (see ``Martian dust climatology'').

In this paper we describe how we made use of newly processed dust opacity retrievals from thermal infrared observations of the MRO/MCS instrument \citep{Mccl:07} in order to reconstruct maps of column dust optical depth specifically for MY 34, and describe aspects of the two-dimensional dust climatology. In Section~\ref{sec:maps}, we discuss the improvements both to the MCS retrievals and to the gridding methodology described in \citet{Mont:15}. In Section~\ref{sec:diuvar} we analyze in general terms the CDOD variability at seasonal scale, and specifically the day-to-day evolution of the GDE and late-winter dust storm. We also address the daily variability observed when reconstructing multiple CDOD maps per sol. In Section~\ref{sec:gcm}, the generated ``MY 34 dust scenario'' (i.e. complete, daily, kriged maps) is used in simulations with the Laboratoire de M\'et\'eorologie Dynamique Mars GCM (LMD-MGCM) to 
\begin{inparaenum} 
\item assess that the impact of the MY 34 GDE on the Martian atmosphere is well accounted for, and
\item verify that the GCM is able to produce some of the diurnal variability observed in the reconstructed multiple CDOD maps per sol, when the dust scenario is only used as initial condition.
\end{inparaenum}
Conclusions are drawn in Section~\ref{sec:conclusions}. Appendix~\ref{sec:appendix} is devoted to clarify the versioning of the CDOD maps.

\section{Building column dust optical depth maps}
\label{sec:maps}

The methodology described in \citet{Mont:15} to grid CDOD values using the IWB, and to spatially interpolate the daily maps using kriging, has been applied to observations by MGS/Thermal Emission Spectrometer (TES), ODY/Thermal Emission Imaging System (THEMIS), and MRO/MCS from MY 24 to MY 32. For MY 33, because of the progressive change in local time of THEMIS observations, the weighting functions of the THEMIS dataset and the MCS dataset were slightly modified to favor the MCS dataset. As is mentioned in the introduction, version 2.0 (v2.1 for MY 33) of both irregularly gridded maps and regularly kriged ones (``dust scenarios'') are available on the MCD project website.

For the specific case of the MY 34 GDE, the MCS team has updated their retrievals of temperature, dust and water ice profiles. We have correspondingly updated the gridding/kriging methodology with the aim of producing a more refined and accurate climatology, both for scientific studies and for the use in numerical model simulations. Therefore, in the following we describe how we reconstruct CDOD maps for MY 34 (currently version 2.5). We provide some details about the differences between current and previous versions (i.e. v2.2, v2.3, and v2.4) in Appendix~\ref{sec:appendix}.

\subsection{Observational dataset}

In MY 34, single THEMIS CDOD retrievals are no longer available. Because of the late local time of THEMIS observations in MY 34, Smith et al. [this issue] had to develop a ``stacking'' algorithm that assesses how a group of THEMIS spectra in a solar longitude/latitude bin change as a function of estimated thermal contrast. Therefore, we do not use THEMIS any more in MY 34 and we completely rely on estimated CDODs from MCS. 

Dust opacity retrievals from thermal infrared observations of the Mars Climate Sounder instrument aboard MRO are described in \citet{Klei:09,Klei:11,Klei:17}. The currently standard MCS dataset, based on the v5.2 ``two-dimensional'' retrieval algorithm specifically described in \citep{Klei:17}, has been reprocessed by the MCS team during the time of the MY 34 GDE to obtain better coverage in the vertical and, therefore, more reliable estimates of CDOD values during the event [Kleinb\"oehl et al., this issue]. This latest MCS dataset, only available between May 21, 2018 ($L_s \sim 179^{\circ}$) and October 15, 2018 ($L_s \sim 269^{\circ}$), labelled v5.3.2, is an interim version that includes the use of a far infrared channel for retrievals of dust. The differences between MCS retrievals version 5.2 and 5.3 are as follows:
\begin{itemize}
    \item Use of B1 detectors to extend the dust profile retrieval: the dust extinction efficiency in channel B1 at 32 $\mu$m is only about half the value of channel A5 at 22 $\mu$m \citep{Klei:17mamo}, which is the primary channel for dust retrievals, allowing profiles to extend deeper by 1 to 1.5 scale heights;
    \item Accepting more opacity from aerosols in the temperature retrieval channel A3;
    \item Modifications for determining surface temperature when there are no matching on-planet views (primarily cross-track views) to improve the performance under high dust conditions when the array is lifted and limb views do not intersect the surface.
\end{itemize}

CDODs are estimated by integrating the dust opacity profiles after an extrapolation from the lowest altitude at which profile information is available, under the assumption of homogeneously mixed dust [see Fig. 1 of Kleinb\"oehl et al., this issue]. For the reconstructed CDOD maps in MY 34, we use MCS v5.3.2 estimated CDODs from $L_s \sim 179^{\circ}$ to $L_s \sim 269^{\circ}$, and MCS v5.2 otherwise.

\subsection{Data Quality Control}
\label{subsec:QC}

A general discussion about the limitations of using CDOD estimates from MCS is included in \citet{Mont:15}, specifically Section~2.1.2. As is mentioned in the previous sub-section, the extended vertical coverage in MCS v5.3.2 helps estimate CDODs more accurately. In the present work, therefore, we have improved the definition of the Quality Control (QC) procedure with respect to the one used in \citet{Mont:15}, particularly by allowing a more extensive use of dayside observations. We define dayside observations as those with local times in the range 09:00 < lt $\leq$ 21:00, although most of dayside observations at low latitudes have local times close to 15:00. Nightside observations are defined as those outside the dayside range, with most nightside observations at low altitude having local times close to 03:00. Note that MCS is also able to observe cross-track, thus providing information in a range of local times at selected positions during the MRO orbits \citep{Klei:13}. We have also better defined a dust quality flag in MCS v5.3.2 to help filtering those observations where a significant number of detectors were excluded in the retrieval of the dust opacity profile, because of radiance residuals exceeding threshold values \citep{Klei:09}. Each excluded detector corresponds to a truncation of $\sim5$~km from the altitude range that was originally selected by the retrieval algorithm based on line-of-sight opacity.

We apply the following QC procedure to the MCS CDOD values at 21.6 $\mu$m in extinction: 
\begin{itemize}
    \item To discard values when they are most likely contaminated by $\mathrm{CO_{2}}$ ice (i.e. if, at any level below 40 km altitude, the temperature is $T < T_\mathrm{{CO_{2}}} + 10 \mathrm{K}$, and the presumed dust opacity is larger than $10^{-5}$~ km$^{-1}$);
    \item To discard values when water ice opacity is greater than dust opacity at the cut-off altitude of the corresponding dust profile;
    \item To discard cross-track CDODs with cut-off altitudes higher than 8 km (i.e. the corresponding dust opacity profiles do not extend down to 8 km altitude or lower);
    \item To discard dayside values with cut-off altitudes higher than 8 km unless it is MCS v5.3.2 data or MCS v5.2 data during the late-winter dust storm (i.e. for $L_s\leq179^{\circ}$, $269^{\circ}<L_s\leq 312^{\circ}$, and $L_s\geq350^{\circ}$). Additionally, we discard dayside CDODs with cut-off altitudes higher than 16 km in a short period before the onset of the GDE ($179^{\circ}<L_s< 186.5^{\circ}$) and during the late-winter dust storm ($312^{\circ}<L_s< 350^{\circ}$). No filtering based on cut-off altitude is applied during the GDE for dayside CDODs;
    \item To discard CDODs when more than 1 detector is excluded inside the limits of the MCS v5.3.2 ($179^{\circ}\leq L_s \leq 269^{\circ}$) as well as if any detector is excluded in MCS v5.2 during the late-winter dust storm ($312^{\circ}<L_s< 350^{\circ}$);
    \item To assign a fixed value of 0.01 to very low values of CDOD < 0.01 having cut-off altitude higher than 4 km.
\end{itemize}

\begin{figure}[ht!] 
\begin{center}
\includegraphics[width=0.9\textwidth]{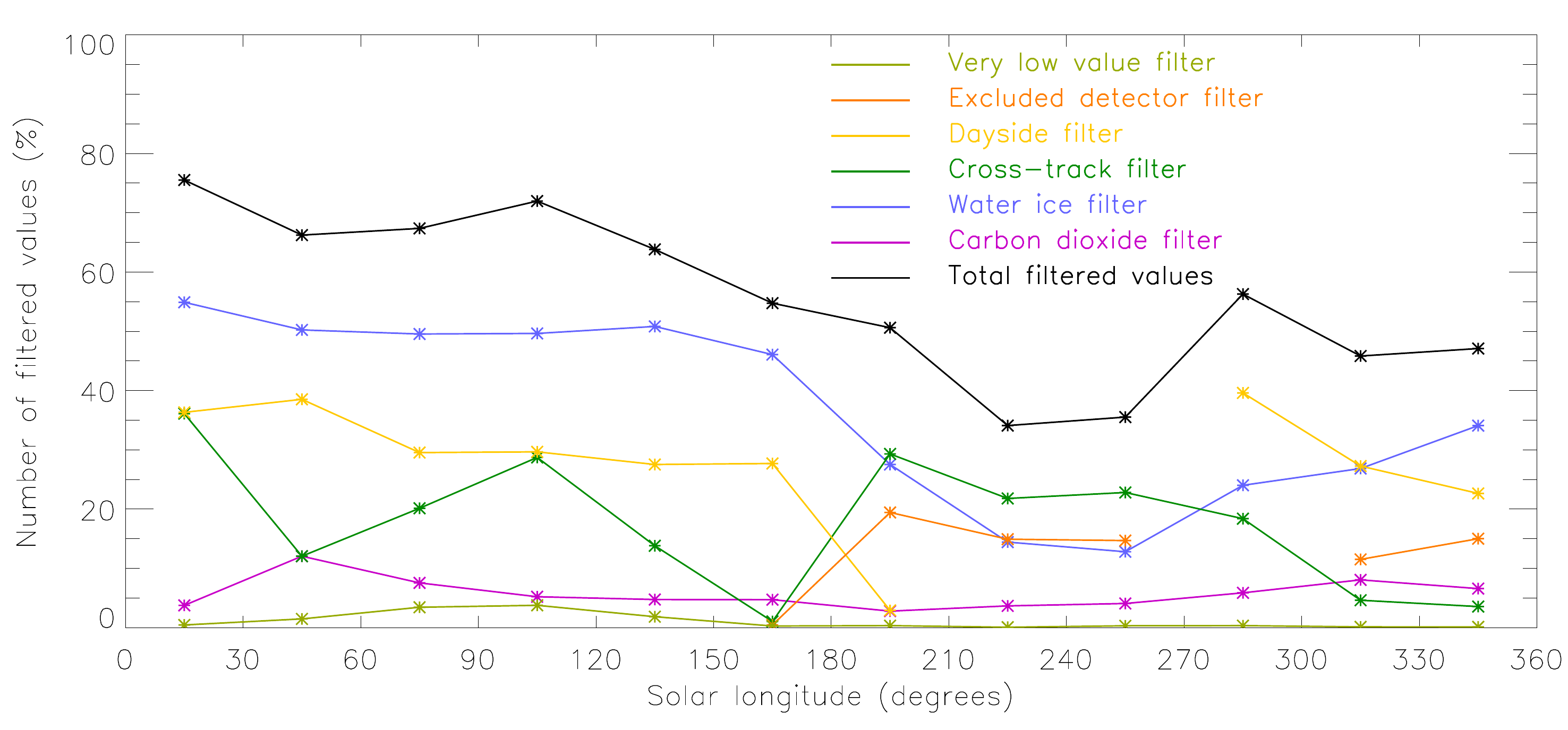}
\caption{\label{fig:QC} \emph{Percentage number of CDOD values flagged by each individual filter in the QC procedure within $30^{\circ}$ solar longitude ranges in MY 34 (color lines), together with the percentage total number of filtered CDODs after the application of the complete QC procedure (black line). The numbers are associated to the middle of each $30^{\circ}$ solar longitude range. Note that the ``excluded detector'' filter does not apply before $L_s=179^{\circ}$ and for $269^{\circ}<L_s< 312^{\circ}$, and that the ``dayside'' filter does not apply during the GDE ($186.5^{\circ} \leq L_s \leq 269^{\circ}$).}}
\end{center} 
\end{figure}

We plot in Fig.\ref{fig:QC} the percentage number of CDOD values that are flagged by each individual filter, together with the total of the filtered values after the application of the complete QC procedure. The total does not correspond to the sum of each single filter, as a CDOD value can be flagged by multiple filters. This figure clearly shows that the presence of water ice in spring and summer strongly affects the number of CDOD values passing the QC. Dayside values are also problematic because their corresponding dust profiles usually have rather high cut-off altitudes, compared to nightside values. Cross-track values have the tendency to exhibit rather high cut-off altitudes as well, and lead to questionable column dust optical depths. As a consequence, a large number of them at low- and mid-latitudes are discarded throughout the year. Observations where more than one detector was excluded in the retrieval are about 20$\%$ throughout MCS v5.3.2. The number of values filtered because of possible carbon dioxide ice contamination is relatively low throughout the year (less than 10\%).

Note that, despite the improvements in MCS v5.3.2, the main issue for estimating CDODs using limb observations remains the fact that many dust opacity profiles have rather high cut-off altitudes (particularly dayside ones, see also third column of Fig.\ref{fig:multiplot}), because of either large dust or water ice opacities. This implies to extrapolate the profiles for a long altitude range under the assumption of homogeneously mixed dust, thus increasing the uncertainty on the column values.  

\begin{figure}[ht!] 
\begin{center}
\includegraphics[width=0.9\textwidth]{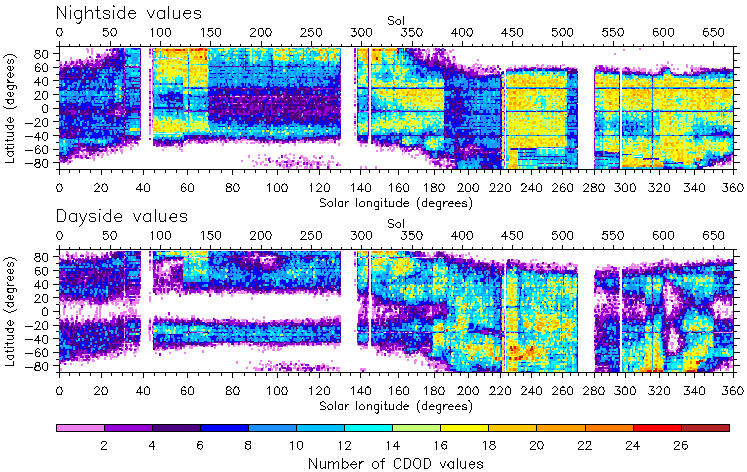}
\caption{\label{fig:nretrievals} \emph{Number of nightside (upper panel) and dayside (lower panel) values of column dust optical depth available for gridding after passing the quality control procedure described in the text. The number of values is summed in 1 sol $\times$ 2$^{\circ}$ latitude bins, and plotted as a function of sol from the beginning of MY 34, corresponding solar longitude, and latitude. Dayside observations are defined to have local times in the range ]09:00, 21:00], whereas nightside observations are defined to have local times in [00:00, 09:00] and ]21:00, 24:00].}}
\end{center} 
\end{figure}

After QC, the number of available values is plotted in Fig.\ref{fig:nretrievals}, separated between nightside and dayside values. The aphelion cloud belt and the winter polar hoods are mainly responsible for the lack of data at equatorial latitudes in the dayside plot, and at high latitudes in both dayside and nightside plots. The implementation of the new ``water ice'' filter is effective in reducing the probability that the lowest levels of the dust profiles are contaminated by the presence of clouds, but there is a risk of filtering out retrievals that actually may have been usable, particularly at high latitudes. A refinement of this filter can be addressed in future work. Vertical bands with no data are periods when MCS did not observe.

\subsection{Data uncertainties and processing}
  
Together with the QC procedure, we have also revised the empirical method to estimate the uncertainties on the MCS CDOD values at 21.6 $\mu$m in extinction, with respect to the one used in \citet{Mont:15}. We apply the following relative uncertainties: 
\begin{itemize}
\item 10\% for CDOD values < 0.01 having cut-off altitude higher than 4 km (i.e. for those values replaced with CDOD=0.01);
\item 5\% for CDOD values < 0.01 or values with cut-off altitudes lower than 4 km;
\item When CDOD $\geq$ 0.01 or cut-off altitude $\geq$ 4 km, we assign the largest relative uncertainty between the one calculated as a linear function of CDOD and the one calculated as a linear function of the cut-off altitude. The two functions are defined in such a way that the uncertainty is 5\% if CDOD = 0.01 or cut-off altitude = 4 km, 20\% if CDOD = 1.5, and 30\% if cut-off altitude = 25 km.


\end{itemize}

As detailed in \citet{Mont:15}, further data processing consists in converting MCS CDODs from 21.6 $\mu$m in extinction to absorption-only 9.3 $\mu$m by multiplying by 2.7, to be consistent with the climatologies of the previous Martian years. We then normalize the values to the reference pressure level of 610 Pa, but instead of using the surface pressure value calculated by the MCD \texttt{pres0} routine \citep{Forg:07omeg}, we now use the same surface pressure value used for the corresponding MCS retrieval. MCS retrieves pressure at the pointing altitude where it is most sensitive to pressure (typically 20-30 km \citep{Klei:09}), from which surface pressure can be extrapolated with an uncertainty estimate based on pointing uncertainty. In conditions where a pressure retrieval is unsuccessful (typically in conditions of high aerosol loading) the MCS algorithm uses pressure derived from the climatological Viking surface pressure \citep{With:2012}. In this case, the uncertainty of the surface pressure is derived from the day-to-day root mean squared of surface pressure from the MCD v5.3, interpolated at the specific location and season of an observation using a pre-built 5$^{\circ}$ solar longitude $\times$ 5$^{\circ}$ latitude array \citep[as described in Section~2.3 of ][]{Mont:15}.

\subsection{Gridding methodology}
\label{subsec:gridmethod}

In this work we closely follow the basic principles of reconstructing CDOD maps, which have been detailed in \citet{Mont:15}. Iterative Weighted Binning (IWB) is applied to CDOD values at 9.3 $\mu$m in absorption, normalized to 610 Pa, together with their corresponding uncertainties, to produce gridded values on a 6$^{\circ}$ longitude $\times$ 5$^{\circ}$ latitude map, with possible ``no value'' assigned at locations where data do not satisfy given requirements. The current criterion to accept a value of weighted average at a particular grid point at any given iteration is that there must be at least one observation within a distance of 200 km from the grid point, while the other parameters listed for MCS in Table~1 of \citet{Mont:15} remain the same.

The key difference with respect to the methodology described in Section~3 of \citet{Mont:15} is that in this work we create four gridded maps per sol at four different Mars Universal Times (MUT, i.e. the local time at 0$^{\circ}$ longitude), opportunely separating the contribution of dayside and nightside observations. We achieve this by 
\begin{inparaenum}
\item considering observations in iterative time windows (TW) centered at four MUTs rather than simply at MUT = 12:00 (this is equivalent to a 6-hour rather than 24-hour moving average), and
\item calculating the local time of each grid point and imposing that only observations with local times within $\pm7$ hours in the current TW are considered for gridding at that grid point.
\end{inparaenum}

\begin{figure}[ph!] 
\begin{center}
\includegraphics[height=10cm,angle=-90]{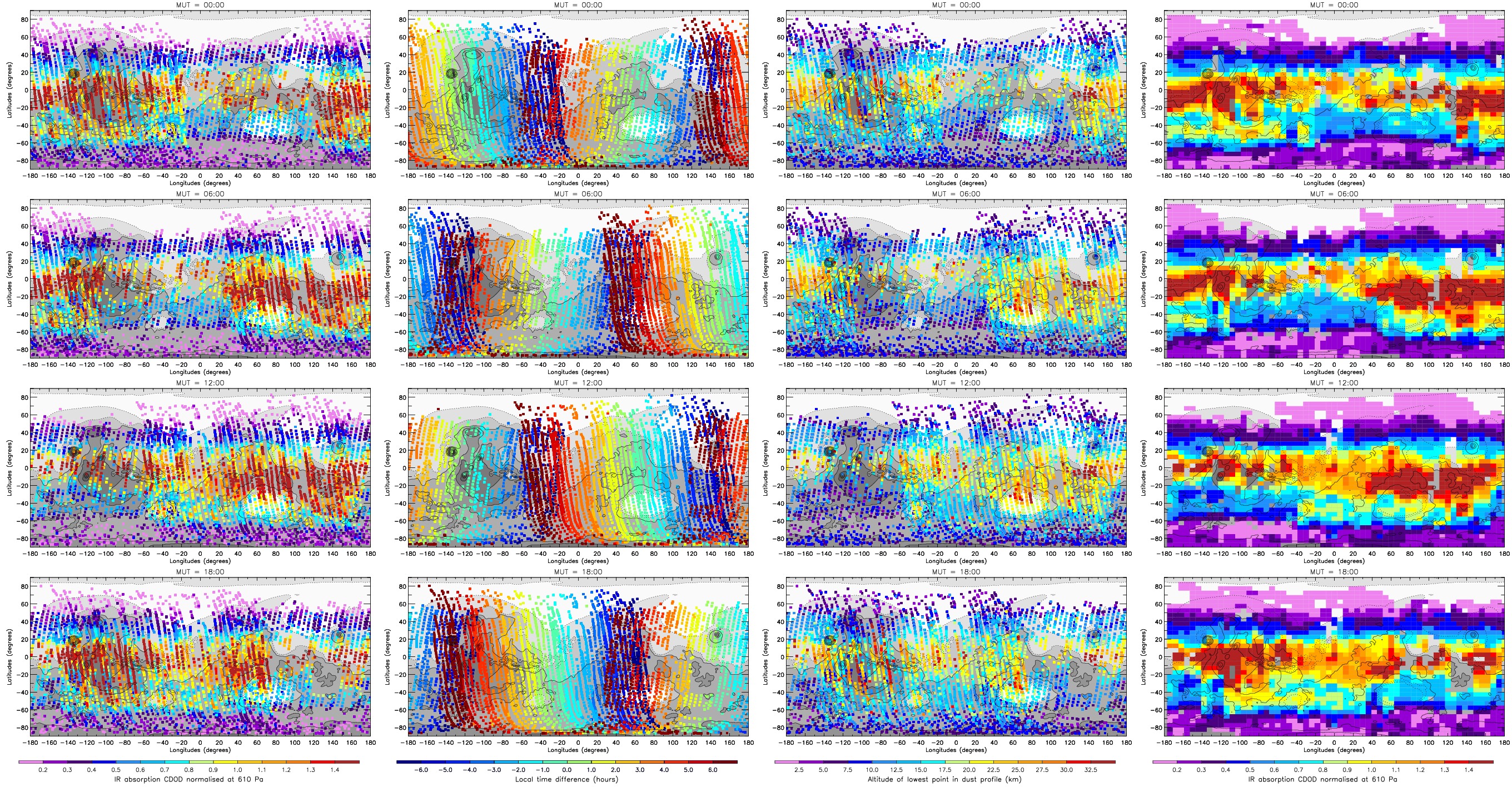}
\caption{\label{fig:multiplot} \emph{This figure shows longitude-latitude maps at four different Mars Universal Times (MUT = 00:00, 06:00, 12:00, and 18:00) in each column, for sol-of-year (SOY) 400, $L_s \sim 196^{\circ}$, in the growing phase of the GDE (The SOY is the integer sol number starting from SOY=1 as first sol of the year). In the first row we plot the values of IR absorption (9.3 $\mu$m) CDOD at 610 Pa for observations within a time window TW = 7 sols centered around the corresponding MUT. In the second row we show the local time difference between the same observations and the grid point around which they are located. The cut-off altitude above the local surface of the dust profile corresponding to each observation is shown in the third column. Finally, the gridded maps of IR absorption (9.3 $\mu$m) CDOD at 610 Pa obtained by applying the iterative weighted binning methodology \citep{Mont:15} to the observations shown in the first three rows are plotted in the fourth row. These gridded maps are already the result of the iterative application of four TW = 1, 3, 5, and 7 sols.}}
\end{center} 
\end{figure}

The result of applying this updated methodology is illustrated as an example in Fig.~\ref{fig:multiplot} for a sol in the growing phase of the GDE. In the first column we show the CDOD values effectively used for gridding (specifically for TW = 7 sols) in the four maps with MUTs = 00:00, 06:00, 12:00, and 18:00. The distinction between nightside tracks (positive slope) and dayside ones (negative slope) is evident. Because we have used a $\pm7$ hour range for accepting observations at each grid point, there is a superposition of nightside and dayside values at some longitudes, which allows for a smoother transition between the two. The distinction in local time is highlighted in the second column of Fig.~\ref{fig:multiplot}. Here we plot the local time difference between each observation and the grid point close to which it is located. Clearly there are two longitude ranges (with local times around 03:00 and 15:00) within which these differences are small, for each different MUT map. These correspond to the locations spanned by the MRO orbit for that specific MUT, whereas higher values of the local time difference correspond to locations spanned by the MRO orbit at a different MUT time, but always within the considered iterative TW. Since we show examples with TW = 7 in Fig.~\ref{fig:multiplot}, there are multiple orbit tracks with similar local time differences, but corresponding to different sols. It is the time weight defined in the IWB procedure, together with the distance and quality of observation weights, that define the contribution of each single observation to the grid point average, plotted in the fourth column. Note that these gridded maps are already the result of the iterative application of four time windows --- refer to Figure~6 of \citet{Mont:15} for an example of how we derive a gridded map using observations in successive TWs.

It is necessary to discuss in depth the differences among the maps at different MUTs, because these are the novel result of this work. When looking at the four MUT maps in the fourth row of Fig.~\ref{fig:multiplot}, in fact, a clear daily variation of CDOD can be appreciated, particularly pronounced in the latitude band [20$^\circ$, 70$^\circ$]S (see also Section~\ref{sec:diuvar} and Fig.~\ref{fig:diuvartimeseries}). This variation of CDOD has the characteristics of a Sun-synchronous wave with wavenumber one: smaller optical depths are found at night, larger optical depths occur during the day. The daily variation is well accounted for in the MCS CDODs, as shown in the first row of Fig.~\ref{fig:multiplot}, and is not an artefact of the gridding methodology, nor is limited to the sol showed in Fig.~\ref{fig:multiplot}, as Fig.~\ref{fig:diuvartimeseries} clearly demonstrates. Furthermore, this strong daily variation of CDOD corresponds very well both in solar longitude (in the growing phase of the GDE) and in latitude to the strong daily variations of the MCS dust opacity profiles, as described in Kleinb\"ohl et al. [this issue. In that paper, GCM simulations can account for the daily variability of the dust profiles, and explain the likely dynamical effects at the origin of this phenomenon. 

The question arises, then, whether the daily variability observed in MCS CDOD can also have a dynamical origin, or can be explained otherwise. We address the possibility of a dynamical origin with GCM simulations in Section~\ref{sec:gcm}, while we point out here that interpreting results from MCS CDODs is particularly challenging, as already mentioned in Subsection~\ref{subsec:QC}. The third row of Fig.~\ref{fig:multiplot}, in fact, shows that the dust profiles in the latitude band where the daily CDOD differences are more pronounced have quite different cut-off altitudes above the local surface between day and night: nightside profiles tend to extend lower in altitude, while dayside profiles are generally cut-off at higher altitudes. This is due to several factors, although it is primarily driven by the altitude at which the retrieval algorithm finds the atmosphere too opaque in the limb path. The increase in the amount of dust (and its vertical extent) in the dayside profiles causes them to terminate further from the surface than the nightside ones, on average.  We have taken into account the quality of the retrieval fit to the measured radiances by selecting profiles with no more than one detector rejected due to fitting poorly. As previously pointed out, the different cut-off altitudes for nightside and dayside retrievals imply that the uncertainty in the CDOD extrapolation is larger during the day, but it does not necessarily imply that the homogeneously mixed dust assumption is not valid, particularly during the peak of the GDE. We refer to Section~\ref{sec:diuvar} for further considerations about this point.

\subsection{Reference MY 34 dust climatology}
\label{subsec:refdatabase}

The gridded and corresponding kriged maps of CDOD described in \citep{Mont:15} have been used as reference multi-annual dust climatology in several studies and applications, including the production of MCD statistics. It is, therefore, compelling to produce a reference MY 34 climatology following the approach established for the previous Martian years.

Although in this work we produce four gridded maps per sol, we calculate the daily average and we use only one map per sol to build the reference MY 34 climatology. We do so because
\begin{inparaenum}
\item the daily variability of MCS CDOD is not yet soundly confirmed by independent observations,
\item it is not clear that using a dust scenario with daily variability in model simulations does not trigger spurious effects, e.g. erroneously forcing the tides, and
\item we would like to be consistent with climatologies from previous Martian years.
\end{inparaenum} 

\begin{figure}[ph!] 
\begin{center}
\includegraphics[width=1.\textwidth]{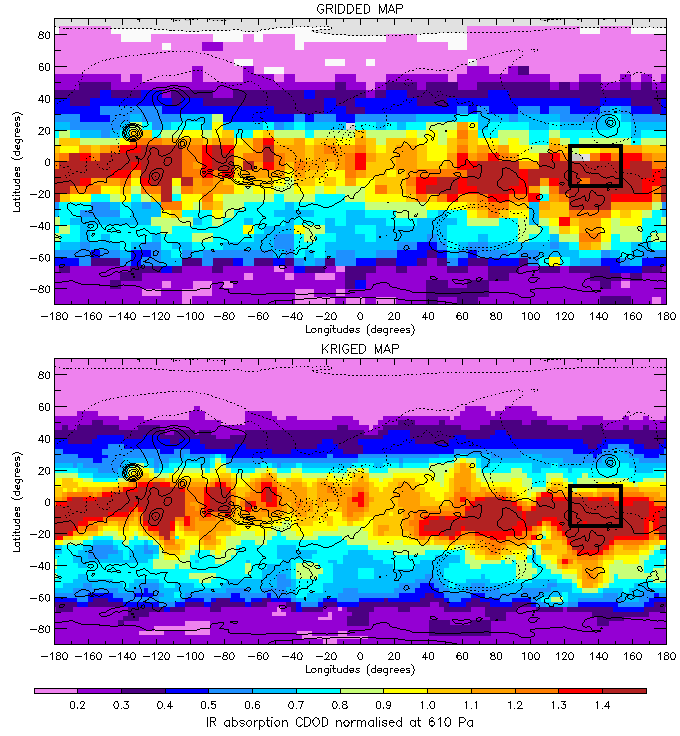}
\caption{\label{fig:gridkrigmaps} \emph{Daily averaged gridded map (upper panel) and corresponding kriged map (lower panel) of 9.3~$\mu$m absorption column dust optical depth for SOY 400, $L_s \sim 196^{\circ}$, in the growing phase of the GDE. The gridded map showed here is the daily average of the four maps in the fourth column of Fig.~\ref{fig:multiplot}. The spatial resolution of the gridded map is 6$^\circ$ longitude $\times$ 5$^\circ$ latitude, whereas the resolution of the kriged one is 3$^\circ$ longitude $\times$ 3$^\circ$ latitude. The black boxes in the maps highlight the averaging area around Gale crater used in  Figures~\ref{fig:validcuriosity} for comparison with the CDOD measured by the Curiosity rover.}}
\end{center} 
\end{figure}

We show in Fig.~\ref{fig:gridkrigmaps} an example of daily averaged gridded map and corresponding kriged one, for the same sol as in Fig.~\ref{fig:multiplot}. The gridded map results more complete than any single MUT map, and rather spatially smooth. The transition to maps at previous and subsequent sols is also rather smooth (see e.g. Figs.~\ref{fig:GDEevolution} and \ref{fig:latestormevolution}). We should mention that, in contrast to \citet{Mont:15}, we no longer modify the values of the gridded maps in a latitude band around the southern polar cap edge before applying the kriging interpolation. This was previously done to artificially introduce climatological ``south cap edge storms'' and balance TES and MCS years in term of dust lifted at the south cap edge. The use of MCS v5.3.2 retrievals extending to lower altitudes, and the fact that TES CDOD retrievals at the south cap edge are being revised (M. Smith, personal communication) alleviates the need for such correction.

The MY 34 daily maps of gridded and kriged IR absorption CDOD normalized to 610 Pa are included in NetCDF files together with maps of several other variables, as detailed in Appendix~B of \citep{Mont:15}. We note that, following the \citet{Mont:15} sol-based Martian calendar (see their Appendix~A for a description), MY 34 has 668 sols, therefore we provide 668 gridded maps --- MY 34 new year's solar longitude is 359.98. The dust scenario, though, has always 669 kriged maps for practical reasons, hence the last sol of the MY 34 dust scenario is the first sol of MY 35. Both gridded and kriged maps version 2.5 for MY 34 will be publicly available at the dedicated webpage on the MCD project website hosted by the LMD at the URL: \url{http://www-mars.lmd.jussieu.fr/} after the version of this manuscript submitted to the Journal of Geophysical Research - Planets is accepted for publication. A beta version of these maps is currently available upon request to the corresponding author.

\subsection{Validation}
\label{subsec:validation}

An important aspect of producing a reference dataset for the dust climatology is its validation with independent observations. The Opportunity rover entered safe mode right at the onset of the GDE, while the Curiosity rover took measurements of visible dust optical depth throughout the GDE using its MastCAM camera \citep{Guze:18}. Hence, we use measurements from Curiosity for validation, together with publicly available visible images taken by the Mars Color Imager (MARCI) camera aboard MRO.

\begin{figure}[ht] 
\begin{center}
\includegraphics[width=1.\textwidth]{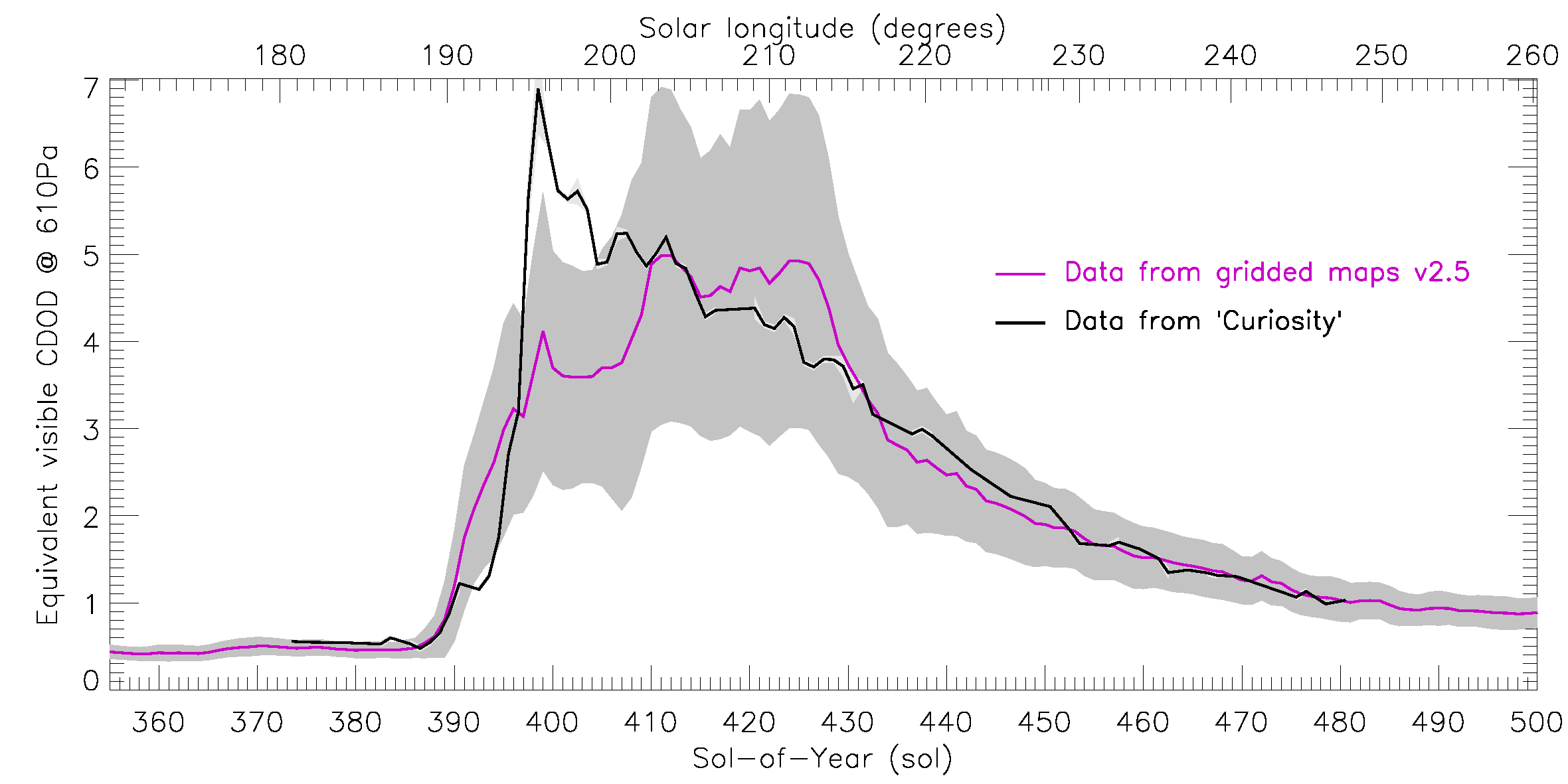}
\caption{\label{fig:validcuriosity} \emph{Time series of equivalent visible column dust optical depth calculated from the 9.3~$\mu$m absorption CDOD normalized to 610 Pa, extracted from the daily averaged gridded maps in an area around Gale crater (magenta line), compared to the time series of visible column optical depth measured by MastCAM aboard NASA's ``Curiosity'' rover (black line). Curiosity observations \citep{Guze:18} have been daily averaged and normalized to 610 Pa (using the surface pressure from the Mars Climate Database \texttt{pres0} routine). Both time series are shown between Sol-of-Year 355 and 500, i.e. $L_s \sim [170^{\circ}, 260^{\circ}]$. We used a factor of 2.6 to convert 9.3~$\mu m$ absorption CDOD into equivalent visible ones. Data from gridded maps are averaged in the area shown by a black box in Fig.~\ref{fig:gridkrigmaps} (i.e. longitude=[123$^{\circ}$E, 153$^{\circ}$E], latitude=[15$^{\circ}$S, 10$^{\circ}$N]) centered around Curiosity landing site at longitude $137.4^{\circ}$E and latitude $4.6^{\circ}$S. Light and dark grey shades show the uncertainty envelope (1-sigma) respectively for Curiosity's time series and the time series extracted from the gridded maps.}}
\end{center} 
\end{figure}

Figure~\ref{fig:validcuriosity} show the comparison between the time series of the dust optical depth (sol-averaged and normalized to 610 Pa) observed by Curiosity rover in Gale crater during the GDE \citep[][]{Guze:18}, and the time series of CDOD extracted from the gridded maps and averaged in a longitude-latitude box centered on Gale crater (after conversion to equivalent visible values). The gridded maps are able to fairly well reproduce the timing and decay of the GDE around Gale, but they underestimate the peak of the event. Furthermore, they overestimate the decay between $L_s \sim 205^{\circ}$ and $L_s \sim 215^{\circ}$, although within the uncertainty limit. There are spatial inhomogeneities in the CDOD field even during the mature phase of the GDE that may account for some of the differences. Also note that Gale crater is a challenging location for MCS to observe due to MRO providing relay services to the Curiosity rover. In particular the number of in-track profiles is limited and may be geographically biased. Finally, if kriged maps were used instead of the gridded maps, the time series would be practically indistinguishable from the magenta line in Fig.~\ref{fig:validcuriosity} (not shown here).

\begin{figure}[ht] 
\begin{center}
\includegraphics[width=1.\textwidth]{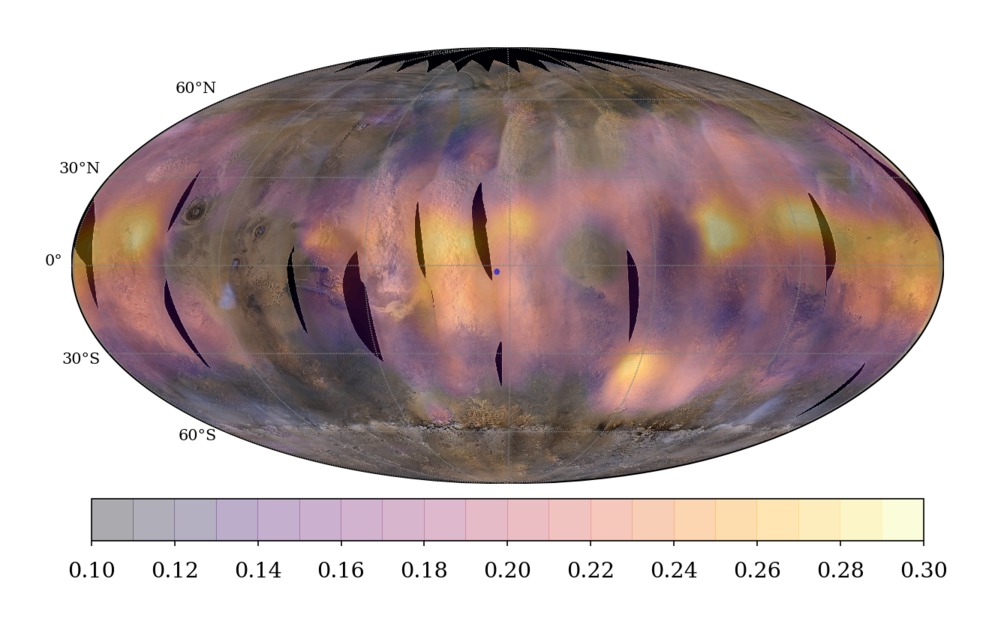}
\caption{\label{fig:validemarci} \emph{The background global image of Mars in this Figure is referenced PIA22329 in the NASA photojournal (credits: NASA/JPL-Caltech/MSSS). It shows the growing MY 34 GDE as of June 6, 2018. The map was produced by the Mars Color Imager (MARCI) camera on NASA's Mars Reconnaissance Orbiter spacecraft. The blue dot shows the approximate location of the Opportunity rover. We overlap on this image the column dust optical depth kriged map for the corresponding sol (sol-of-year 387), which we have reconstructed from MCS observations. The IR absorption (9.3 $\mu$m) CDOD map (not normalized to 610 Pa) is plotted as filled colored contours.}}
\end{center} 
\end{figure}

We show the comparison between one of our gridded CDOD maps and a MARCI image in Fig.~\ref{fig:validemarci}. The comparison is done for June 6, 2018, at the onset of the GDE, which corresponds to SOY 387 in our dataset. The extension of the dust cloud in both the MARCI image and the CDOD map is similar, with both showing intense activity around Meridiani, an eastward progression of the storm, and relatively clear skies over the Tharsis volcanoes. This specific CDOD map fails to show the onset of the south polar cap edge dust activity, but maps at subsequent sols do.

\section{Seasonal, day-to-day, and daily variability of column dust \label{sec:diuvar}}

\begin{figure}[ht]
\begin{center}
\includegraphics[width=1.\textwidth]{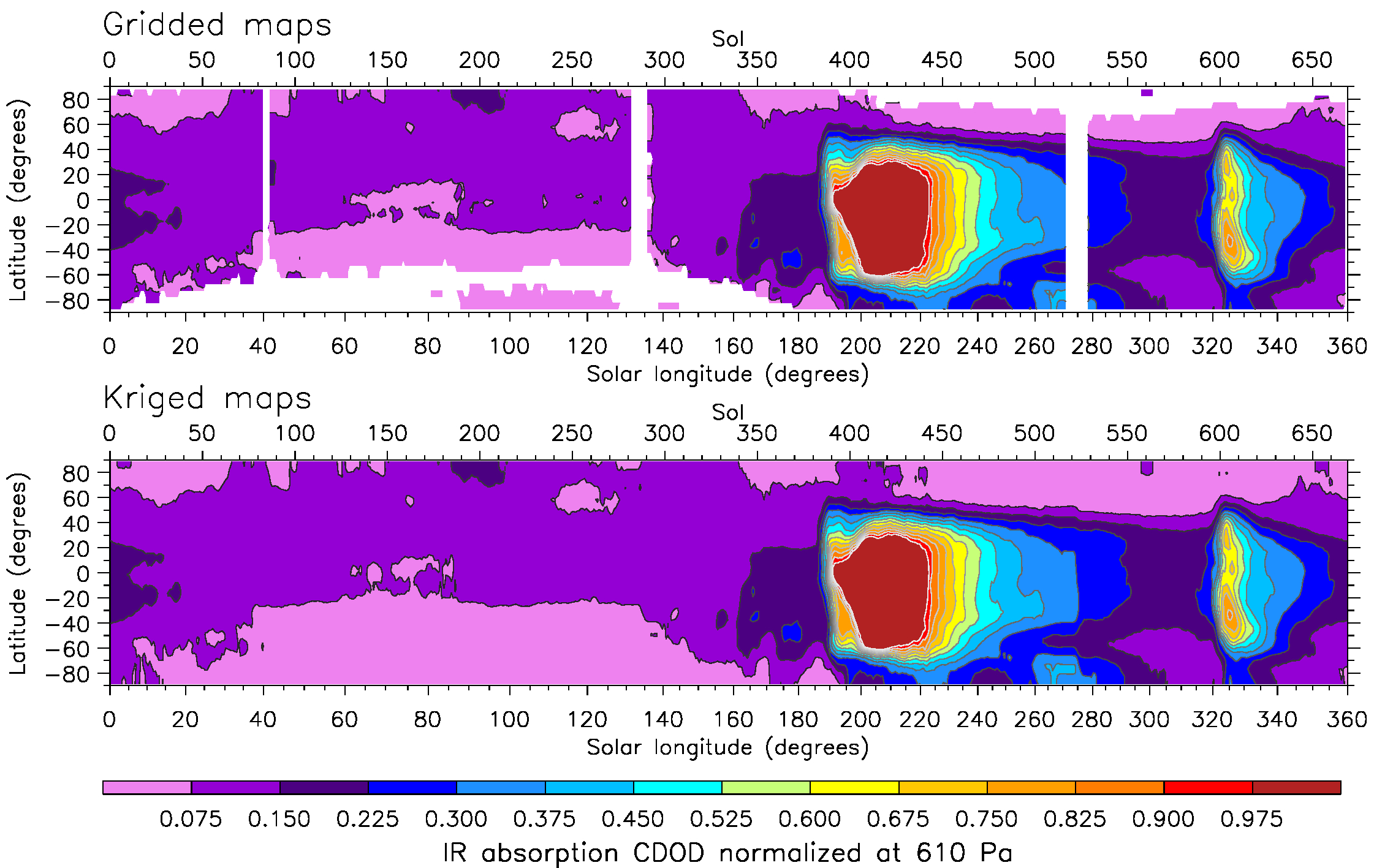}
\caption{\label{fig:zonalmeans} \emph{MY 34 latitude vs time plot of the zonally and daily averaged gridded maps of 9.3 $\mu$m absorption column dust optical depth normalized to the reference pressure level of 610 Pa (upper panel), compared to the same using kriged maps (lower panel). White colour in the upper panel indicates missing data.}}
\end{center}
\end{figure}

\begin{figure}[ht]
\begin{center}
\includegraphics[width=1.\textwidth]{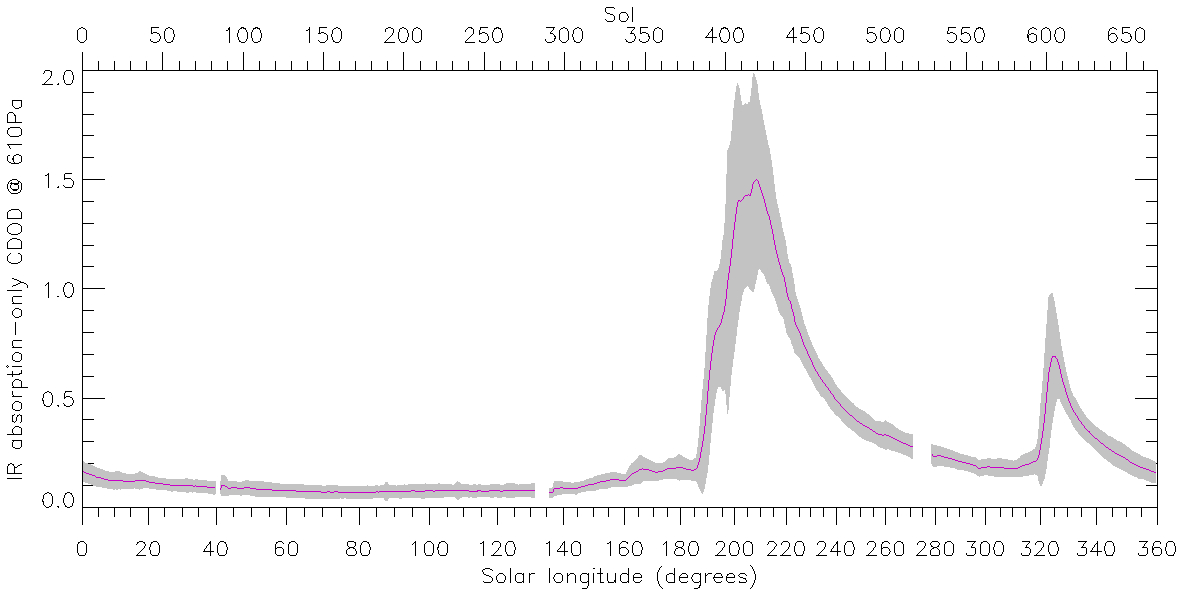}
\caption{\label{fig:CDODtimeseries} \emph{Time series of column dust optical depth (9.3~$\mu$m in absorption, normalized to 610 Pa) extracted from the daily averaged gridded maps and averaged at all longitude in the latitude band [60$^{\circ}$S, 40$^{\circ}$N]. The grey shade represents the root mean squared deviation, i.e. the spatial variability within the averaged longitudes and latitudes (note that the daily variability is not included).}}
\end{center}
\end{figure}

\begin{figure}[hp!]
\begin{center}
\includegraphics[width=0.85\textwidth]{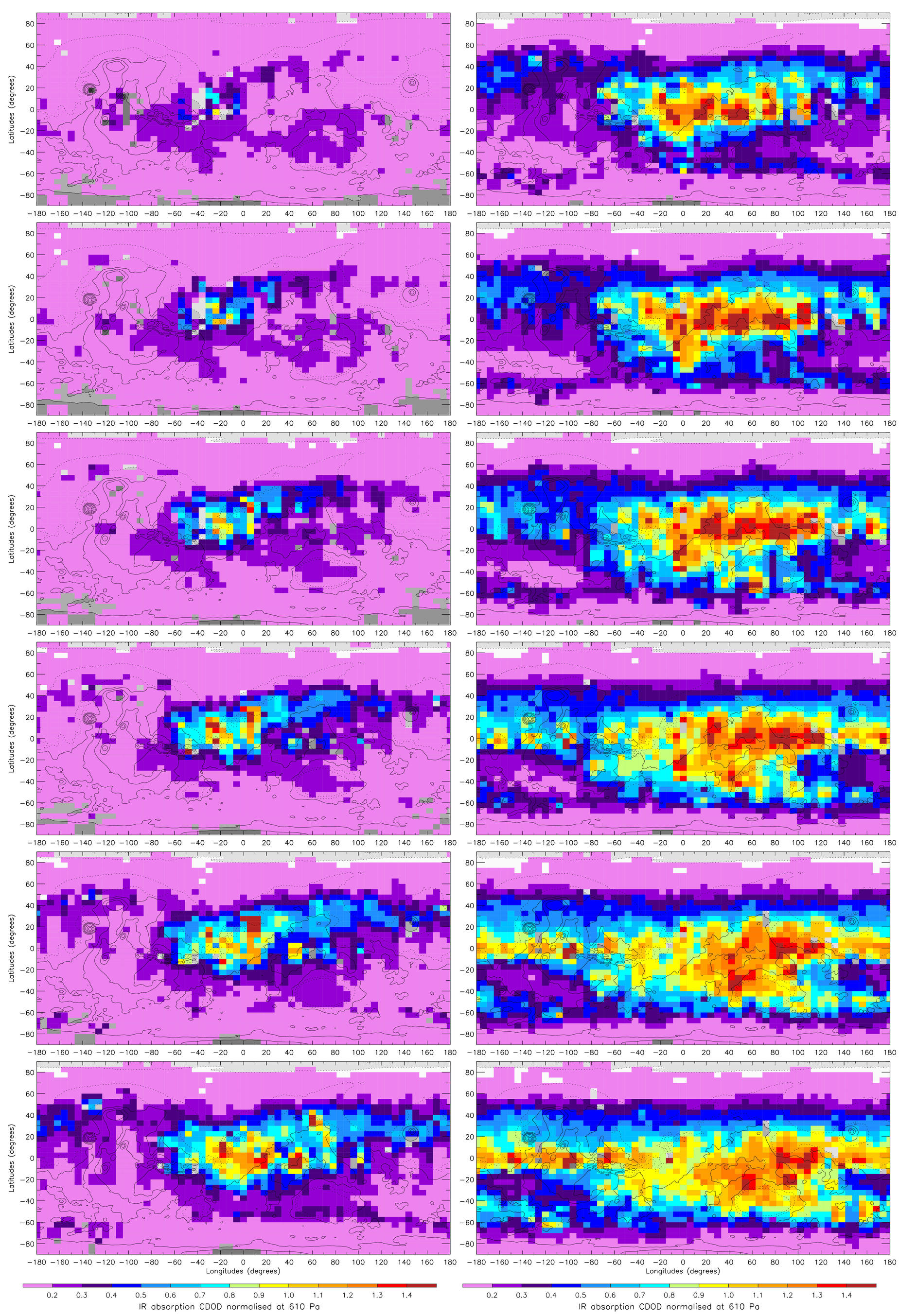}
\caption{\label{fig:GDEevolution} \emph{Initial evolution of the MY 34 Global Dust Event. Each panel shows daily averaged gridded column dust optical depth (in absorption at 9.3$\mu$m) normalized to the reference pressure level of 610 Pa. From top left to bottom right, maps are provided from sol-of-year 383 to sol-of-year 394 (i.e. from $L_s \sim 186^\circ$ to $L_s \sim 193^\circ$). See also Appendix~A of \citet{Mont:15} for the description of the sol-based Martian calendar we use in this paper.}}
\end{center}
\end{figure}

\begin{figure}[hp!]
\begin{center}
\includegraphics[width=0.85\textwidth]{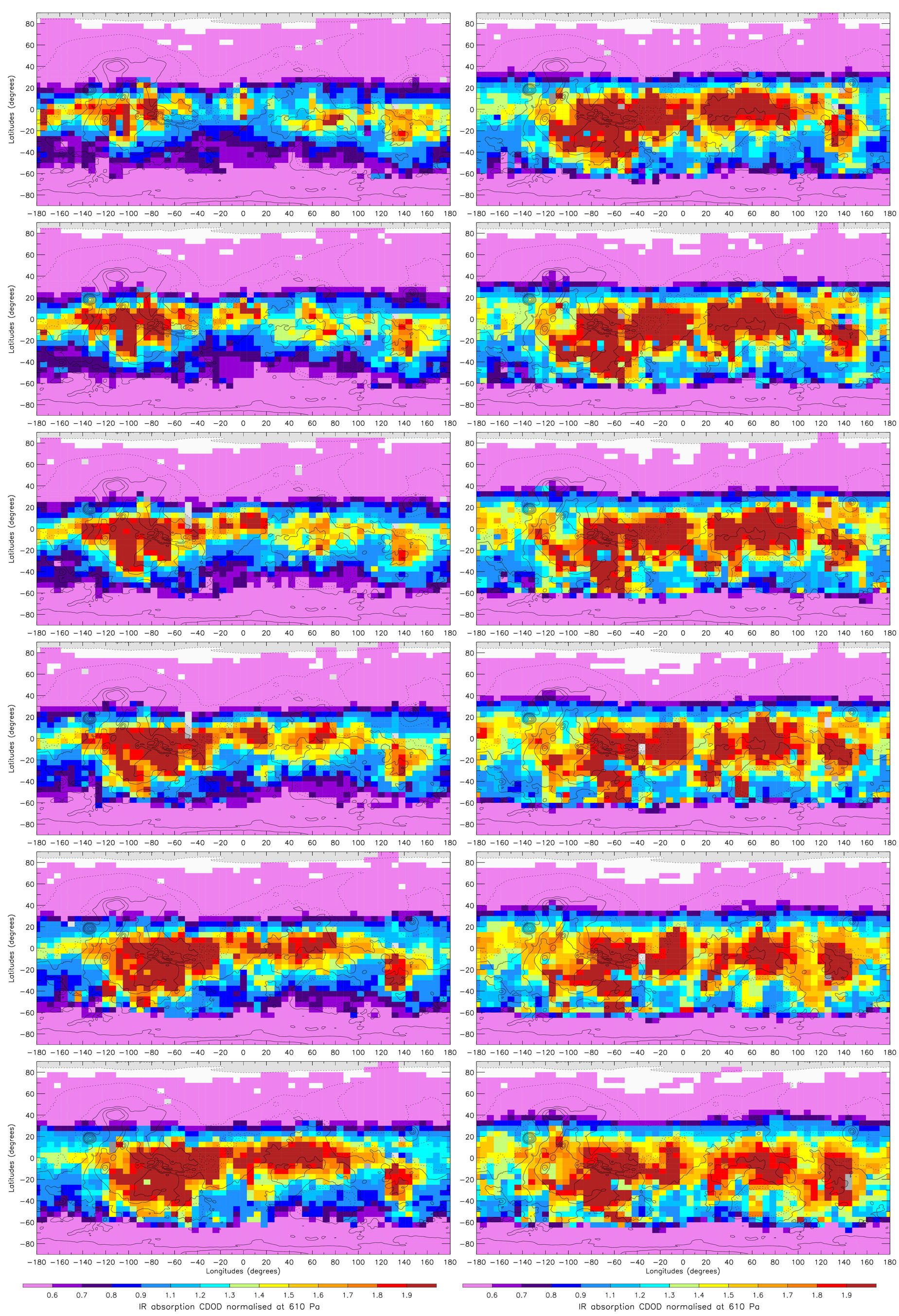}
\caption{\label{fig:GDEsecondevolution} \emph{Same as Fig.\ref{fig:GDEevolution} but for the MY 34 secondary storm within the GDE. From top left to bottom right, maps are provided from sol-of-year 401 to sol-of-year 412 (i.e. from $L_s \sim 197^\circ$ to $L_s \sim 204^\circ$). Note that the scale for the CDOD values has changed with respect to Fig.~\ref{fig:GDEevolution}.}}
\end{center}
\end{figure}

\begin{figure}[hp!]
\begin{center}
\includegraphics[width=0.85\textwidth]{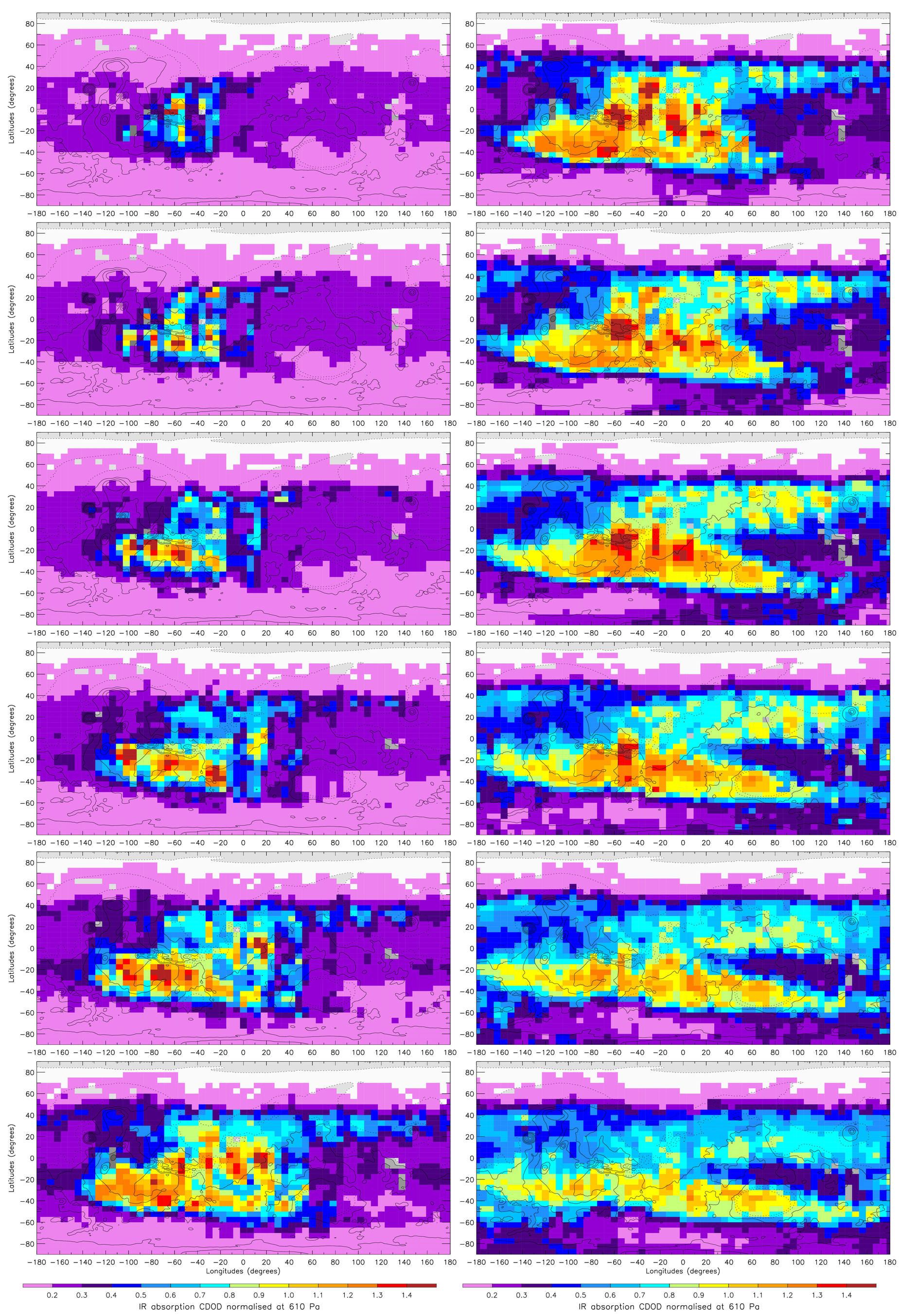}
\caption{\label{fig:latestormevolution} \emph{Same as Fig.\ref{fig:GDEevolution} but for the MY 34 late-winter regional storm. From top left to bottom right, maps are provided from sol-of-year 596 to sol-of-year 607 (i.e. from $L_s \sim 320^\circ$ to $L_s \sim 326^\circ$). The scale for the CDOD value is the same as in Fig.~\ref{fig:GDEevolution}.}}
\end{center}
\end{figure}

\begin{figure}[ht]
\begin{center}
\includegraphics[width=1.0\textwidth]{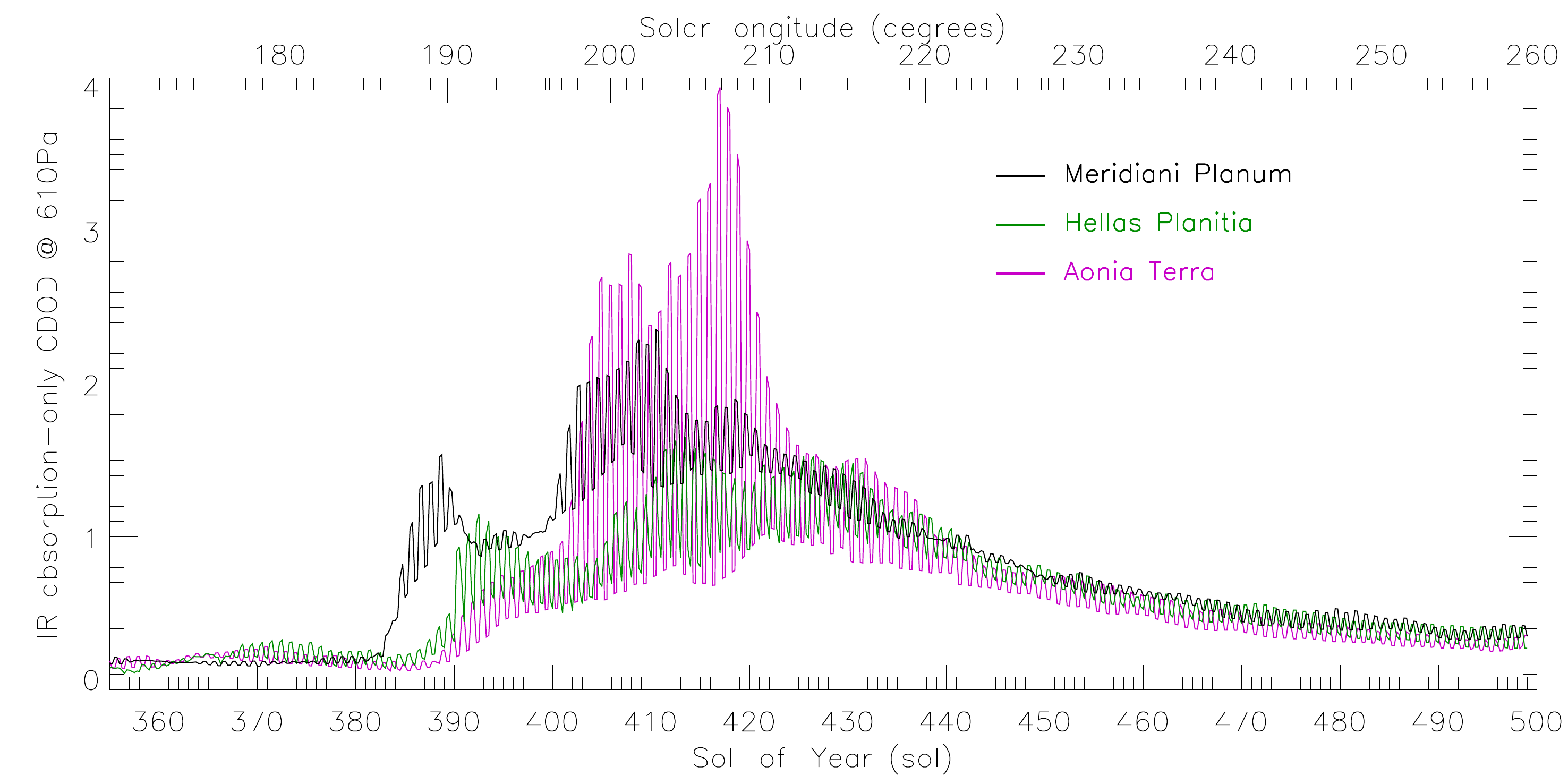}
\caption{\label{fig:diuvartimeseries} \emph{Time series of column dust optical depth (9.3~$\mu$m in absorption, normalized to 610 Pa) extracted from the gridded maps with four MUT per sol in three different areas: Meridiani Planum (longitude=[15$^{\circ}$W, 15$^{\circ}$E], latitude=[15$^{\circ}$S, 15$^{\circ}$N), Hellas Planitia (longitude=[55$^{\circ}$E, 85$^{\circ}$E], latitude=[60$^{\circ}$S, 30$^{\circ}$S), and Aonia Terra (East of Argyre Planitia: longitude=[90$^{\circ}$W, 60$^{\circ}$W], latitude=[60$^{\circ}$S, 30$^{\circ}$S). The time series are shown between Sol-of-Year 355 and 500, i.e. $L_s \sim [170^{\circ}, 260^{\circ}]$.}}
\end{center}
\end{figure}

\begin{figure}[ht!]
\begin{center}
\includegraphics[width=0.9\textwidth]{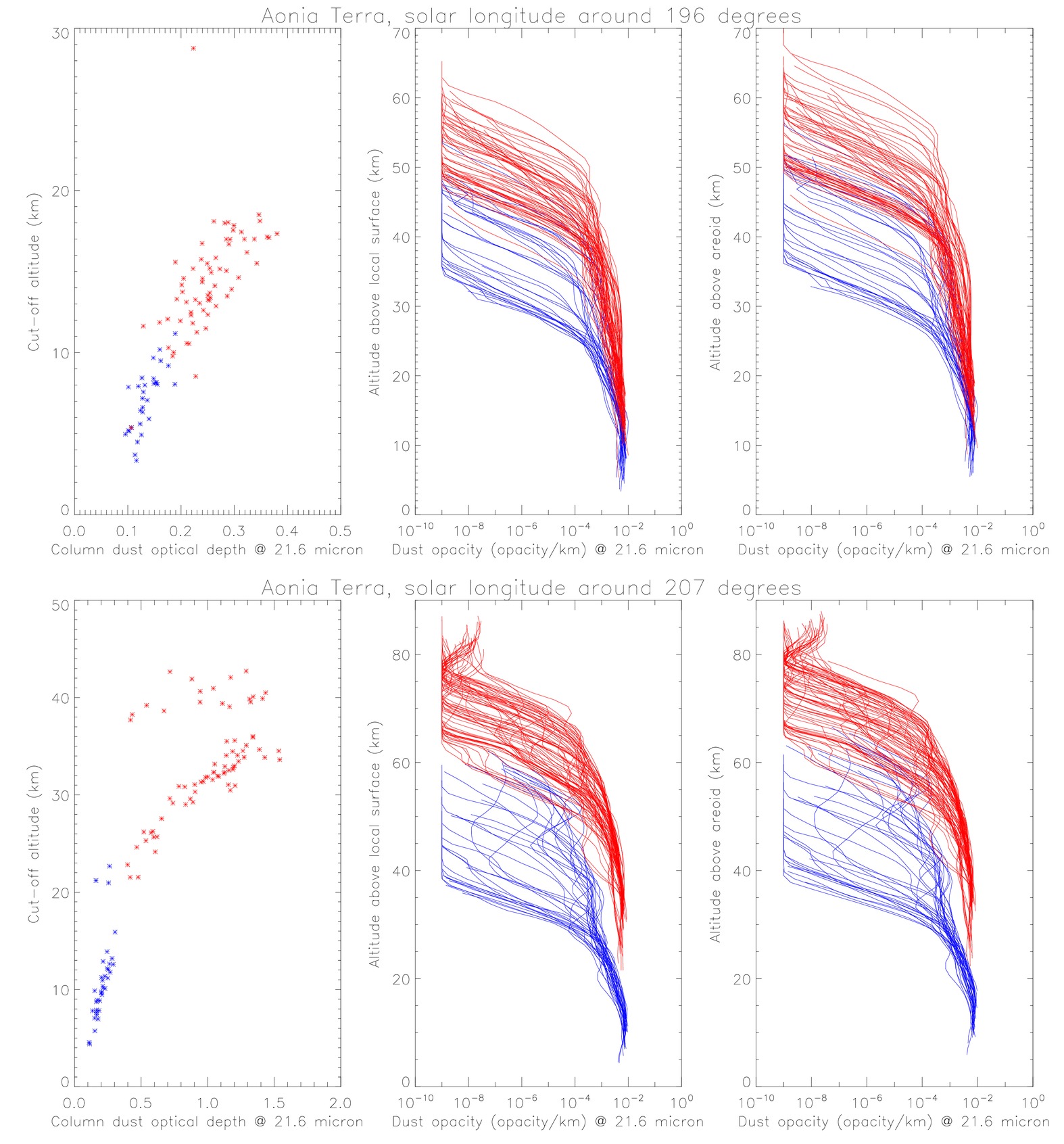}
\caption{\label{fig:dustprofiles} \emph{Data plotted in each panel of this figure are for 7 sols centered on either sol-of-year 400 ($L_s \sim [196^{\circ}$, upper panels) or sol-of-year 418 ($L_s \sim [207^{\circ}$, lower panels) in Aonia Terra (longitude=[90$^{\circ}$W, 60$^{\circ}$W], latitude=[60$^{\circ}$S, 30$^{\circ}$S). Blue indicate nightside data, red is for dayside data. The left panels of this figure shows the MCS CDOD in extinction at 21.6 $\mu$m (not normalized to 610 Pa) as a function of the cut-off altitudes of their corresponding dust opacity profiles, which are plotted in the central panels as a function of altitude above local surface, and in the right panels as a function of altitude above the areoid (topography values are interpolated from the MOLA dataset at the corresponding longitudes and latitudes). Note that the x-axis and y-axis ranges are different among the panels.}}
\end{center}
\end{figure}

In this Section we analyze the variability at different temporal scales, which is included in the MY 34 dust climatology reconstructed from MCS CDODs. In particular, we look at the seasonal, day-to-day, and daily variability, as showed in Figures~\ref{fig:zonalmeans} to \ref{fig:dustprofiles}. 

Starting from the seasonal variability, Fig.~\ref{fig:zonalmeans} shows the latitude vs time plot of the zonally and daily averaged CDOD obtained from both the gridded maps and the kriged ones. This comparison shows that the kriged maps have the advantage of being complete (i.e. CDOD values are assigned at every grid point) while preserving the overall properties of the dust distribution. \citet{Mont:18book} noted that Martian years show two distinctive seasons with respect to the atmospheric dust loading, when a comparison of multi-annual zonal means of CDOD is carried out: a ``low dust loading'' (LDL) season between $L_s \sim 10^\circ$ and $L_s \sim 140^\circ$, and a ``high dust loading'' (HDL) season at other times, when regional dust storms and global dust events are most likely to occur --- commonly referred to as the ``dust storm season''. MY 34 does not differ, as dust started to increase above the 0.15 level (IR absorption at 9.3 $\mu$m) after $L_s \sim 160^\circ$, following a quiet LDL season (see Fig.~\ref{fig:CDODtimeseries} as well, which is the time series obtained from the latitude vs time plot by averaging also in the latitude band [60$^\circ$S, 40$^\circ$N]). 

Nevertheless, the optical depth abruptly increased after $L_s \sim 186^\circ$ due to the onset of the GDE, which rapidly grew to the west of Meridiani Planum, expanded eastwards and southwards, and spread a large amount of dust at all longitudes within approximately a latitude band [60$^\circ$S, 40$^\circ$N] (see its day-to-day evolution over 12 sols in Fig.~\ref{fig:GDEevolution}), then slowly decaying over about 130 sols, as can be observed from the tail of the GDE peak in Fig.~\ref{fig:CDODtimeseries}. 

MY 34 also featured two other maxima in CDOD that are climatologically consistent with all other 10 previously observed years: one at southern polar latitudes centered at $L_s \sim 270^\circ$, and the other one in a latitude band [60$^\circ$S, 40$^\circ$N] peaking at $L_s \sim 325^\circ$. These maxima are linked to a regional dust storm occurring over the ice-freed southern polar region, and to a particularly intense late-winter regional storm (see its day-to-day evolution over 12 sols in Fig.\ref{fig:latestormevolution}). The latter has the characteristics of a flushing storm following the Acidalia-Chryse storm track, although its precise origin cannot be easily tracked in the gridded maps of CDOD. Finally, the absence of significant storms in a range of solar longitude centered around $L_s \sim 300^\circ$ --- the so-called ``solsticial pause'' --- is also climatologically consistent with what observed in previous years \citep[][and Xiaohua et al., this issue]{Mont:15,Kass:16,Lewi:16,Mont:18book}.

Before moving to the analysis of the CDOD daily variability, there is a final consideration we ought to do about the variability of dust storms. When comparing the day-to-day evolution of the GDE and the late winter storm in Figs.~\ref{fig:GDEevolution} and \ref{fig:latestormevolution} at their early stage, they look pretty similar both in intensity and extension. Furthermore, the shapes of the CDOD peaks in the time series of Fig.~\ref{fig:CDODtimeseries} are also comparable (both positively skewed, with sharp increase and long decreasing tail). What really does make the difference is the fact that a GDE such as the one in MY 34 took about 35 sols of continuous dust injection into the atmosphere to reach a peak in average CDOD that is more than twice as high than the one reached by the (still rather intense) late-winter regional storm. This includes a much larger spatial variability during the GDE, as indicated by the RMSD in Fig.~\ref{fig:CDODtimeseries}. An event that was very important in boosting the equinoctial dust storm into the GDE class was the activation of secondary lifting centers in the Tharsis region, which seems to start around SOY 401 in the gridded maps ($L_s \sim 197^\circ$), and later in the Terra Sabaea region --- although one cannot distinguish from the maps whether the increase of optical depth in this region is the result of eastward transport from Tharsis or local dust lifting, or both. When looking at the CDOD day-by-day evolution in Fig.~\ref{fig:GDEsecondevolution}, this event can be considered as a second dust storm within the first one, without which we could have just witnessed a regional storm instead of a GDE. This is one of the reasons why names like ``global dust storm'' or ``planet-encircling dust storm'' do not seem to capture the real nature of this type of extreme events, which are not single storms nor uniquely planet-encircling. Perhaps even ``global dust event'' is not particularly appropriate, as high latitude regions are mostly free of dust --- although indirectly affected by the dust via dynamical effects. One possibility is to call these events for what they really are: ``extreme dust events'' (EDE). 

Another extreme characteristic of the equinoctial event is its strong daily variability, clearly observed by MCS in the elevation of the dayside vs nightside dust profiles (see Kleinb\"ohl et al, this issue), but also featured in the column optical depth values, as already shown in Fig.~\ref{fig:multiplot}. The time series at different locations extracted from the dataset with four MUT maps per sol and shown in Fig.~\ref{fig:diuvartimeseries} clearly illustrates this phenomenon. The nightside-dayside variability is different at different locations, but is particularly dramatic in Aonia Terra, which is located in the southern latitude band where Kleinb\"ohl et al [this issue] observe strong variability in the dust profiles. In Fig.~\ref{fig:dustprofiles}, therefore, we compare the CDOD values in Aonia Terra at two times during the GDE (i.e. during its growth phase and near the peak) with the corresponding dust opacity profiles that are extrapolated and integrated in order to estimate the CDODs. The raise in altitude of about 20 km (either above local surface or above the areoid) of the dayside dust profiles with respect to the night profiles is quite spectacular at $L_s \sim 207^{\circ}$, near the peak of the GDE. Unfortunately, with the raise in altitude of the large opacity values comes the raise in cut-off altitude of the dust profiles. However, from Fig.~\ref{fig:dustprofiles} one cannot conclude that the homogeneously mixed dust hypothesis at the core of our dust profile extrapolation to the ground does not hold in these cases. Conversely, there is no evidence that raising dust is replaced by more well mixed dust in the missing part of the profile, because we simply have no data. Furthermore, uncertainties at the lowest levels of dust profiles during the GDE tend to be larger (see e.g. left panel of Fig.~1 in Kleinb\"ohl et al. this issue), hence the real shape of the profile in the lowest two scale heights could provide some surprise.

At this point of the analysis, we can make three hypotheses about the daily variability observed in MCS CDOD:
\begin{enumerate}
    \item There is an intrinsic variability of the column dust. In this case, quite a substantial amount of dust must be supplied in the lowest two scale heights during the day, where MCS cannot see through. This extra dust must either be lifted locally from the ground or supplied via horizontal advection (or both). Local mesoscale effects might operate at different locations (e.g. katabatic/anabatic winds, strong convective activity, etc.)
    \item There is no significant variability in the column dust. In this case, atmospheric dust is simply moved up and down during the day/night, and the dayside dust opacities actually decrease with decreasing altitude in the lowest scale heights, not seen by MCS. The daily variability of the dust opacity profiles in the lowest two scale heights during the GDE would then be expected to be very large, in order to compensate for the raise in altitude of the dust cloud.
    \item There is some variability in the column dust. In this case, dust is partly moved up and down at different local times, and partly lifted locally, or advected from nearby locations. 
\end{enumerate}
In order to help us clarifying which hypothesis is more likely, we have carried out simulations with the LMD-GCM, which we are going to discuss in the next section.

\section{Global Climate Model simulations of the MY 34 Global Dust Event \label{sec:gcm}}

The simulations we have carried out using the LMD-MGCM have similar characteristics to those carried out to build the Mars Climate Database version 5.3 \citep{Mill:15}, except for the model top being set at 100~km (with 29 vertical levels) and the thermospheric parameterizations \citep{Gonz:11} being switched off. The most up-to-date physical parameterizations are included: interactive dust cycle \citep{Made:11}, thermal plume model \citep{Cola:13}, water cycle with radiative effect of clouds \citep{Made:12radclouds} and full microphysics scheme \citep{Nava:14jgr}. The ``rocket dust storm'' parameterization recently built and tested by \cite{Wang:18} is not included in this version of the GCM. The horizontal grid features~$64 \times 48$ longitude-latitude points. 

The initial state for the MY 34 run at $L_s = 0^{\circ}$ uses the ``climatological'' dust scenario typical of MYs devoid of global dust events. Then, two types of simulations are carried out:
\begin{enumerate} 
    \item a simulation using the reference MY 34 dust scenario v2.5 (i.e. the maps kriged from the daily averaged gridded maps, as discussed in Section~\ref{subsec:refdatabase}) to force the dust column field throughout the GDE period;
    \item a simulation using the MY 34 dust scenario until $L_s = 210^{\circ}$ (around the peak of the GDE), then continuing as a ``free-dust'' run for a few sols, with no more external forcing on the column dust field.
\end{enumerate}
Furthermore, given the availability of a dust scenario for MY 25 \citep{Mont:15} and the presence of an equinoctial GDE in that year, with comparable characteristics to the one in MY 34, we have also carried out a forced simulation for MY 25, for comparison.

The LMD-MGCM simulations for MY 34 are also used in Kleinb\"ohl et al. [this issue] to discuss the diurnal cycle of the vertical distribution of dust observed by MCS. We use the two types of simulations for two different purposes:
\begin{inparaenum}
\item the forced run is used to analyse some of the impacts of the MY 34 GDE on the local and global dynamics, thus verifying that the use of a daily averaged dust scenario produces reasonable results, consistent with those for MY 25;
\item the ``free dust'' run is used to identify possible daily variability of the column dust in the model, which could corroborate one of the three hypotheses provided at the end of last Section,
\end{inparaenum}

\begin{figure}[ht] 
\begin{center}
\includegraphics[width=0.9\textwidth]{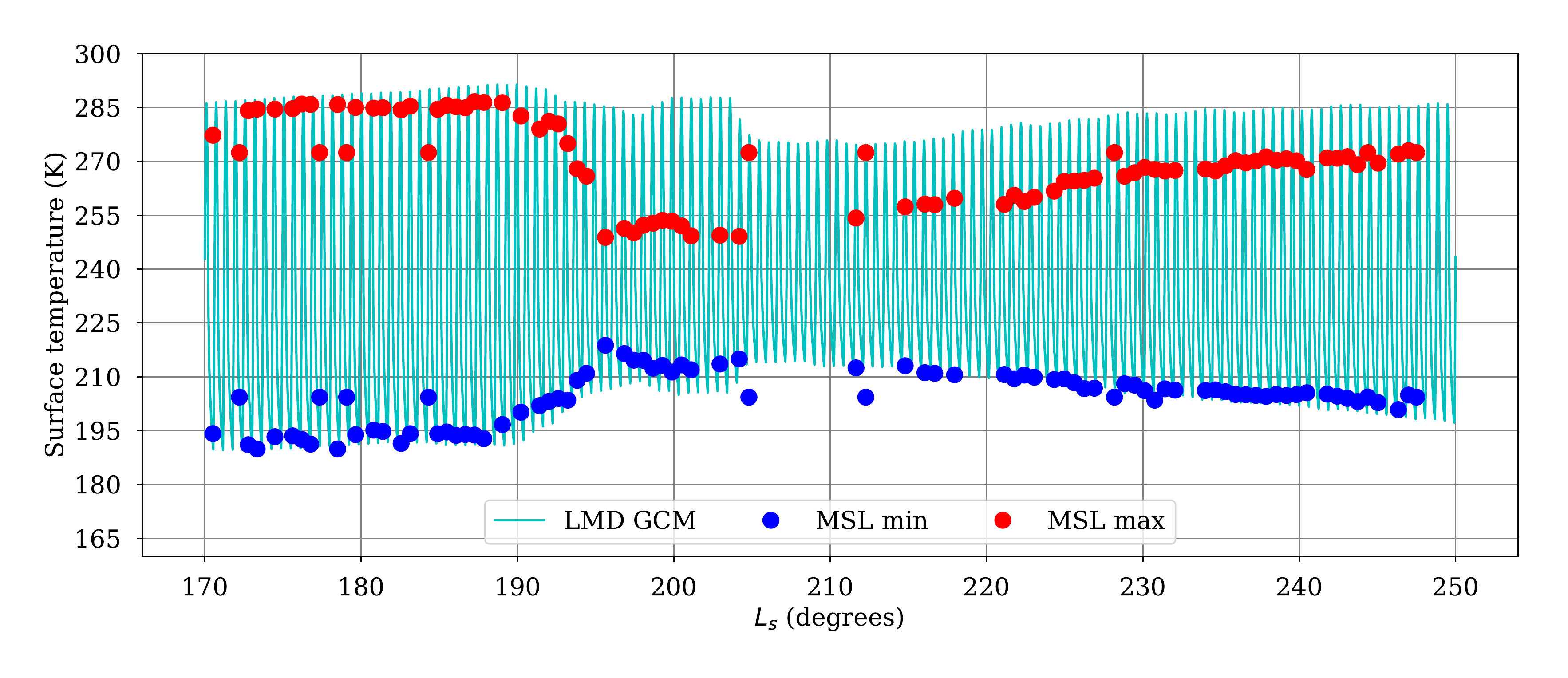}
\caption{\label{fig:seasonmsl} 
\emph{Comparison of surface temperature simulated by the LMD-MGCM model versus surface temperature measured by the Rover Environmental Monitoring Station on board MSL ``Curiosity'' rover (daily minimum in blue and daily maximum in red). Data from MSL are provided as supplementary material of \citet{Guze:18}}}
\end{center} 
\end{figure}

Figure~\ref{fig:seasonmsl} shows a comparison between the surface temperature measured by the Mars Science Laboratory \citep[MSL Curiosity][]{Guze:18} and the surface temperature computed by the LMD Mars GCM. When the MY 34 global dust event starts, the diurnal amplitude of temperature is reduced: daytime temperatures are lower as a result of visible absorption of incoming sunlight being more efficient in a dustier atmosphere, and nighttime temperatures are higher as a result of increased infrared radiation emitted towards the surface in a dustier atmosphere. The temporal variability of temperature (absolute and relative values) is well reproduced for the nighttime minimum temperature, but less so for the daytime temperatures (although the qualitative behaviour is correct). There might be three reasons for this:
\begin{inparaenum}
\item thermal inertia is not well represented in the LMD-MGCM for daytime conditions in Gale Crater;
\item the CDOD observed by MCS in the region of Gale crater is underestimated with respect to the one observed by Curiosity from $L_s \sim 195^{\circ}$ to $L_s \sim 202^{\circ}$, and by consequence the corresponding gridded maps and dust scenario are low-biased at those times (see Fig.~\ref{fig:validcuriosity} in Section~\ref{subsec:validation});
\item the accuracy of the calculations by the model radiative transfer could decrease under extreme dust loading conditions, or could be affected by an inaccurate distribution of particle sizes.
\end{inparaenum}

\begin{figure}[h!] 
\begin{center}
\includegraphics[width=0.45\textwidth]{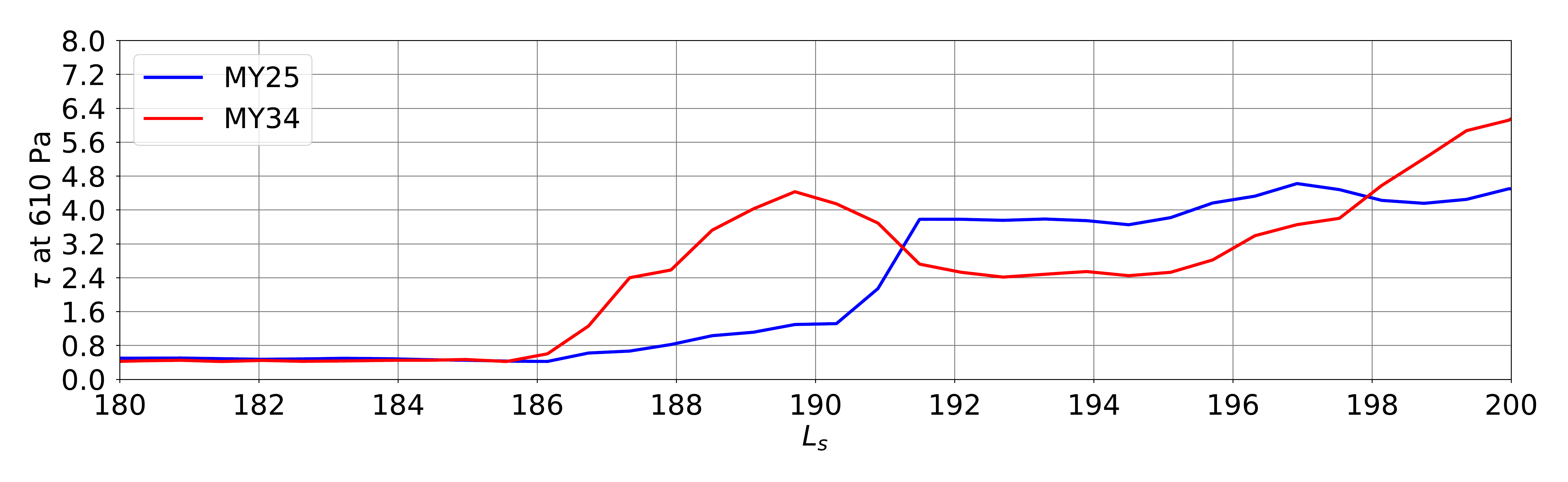}
\includegraphics[width=0.45\textwidth]{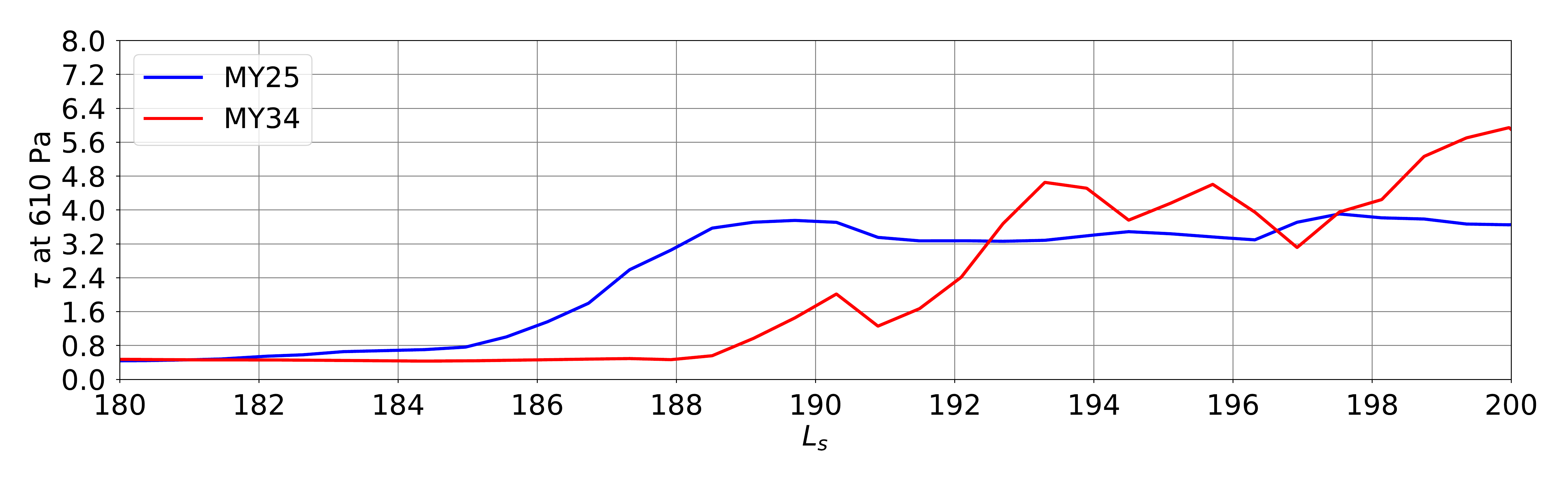}
\includegraphics[width=0.45\textwidth]{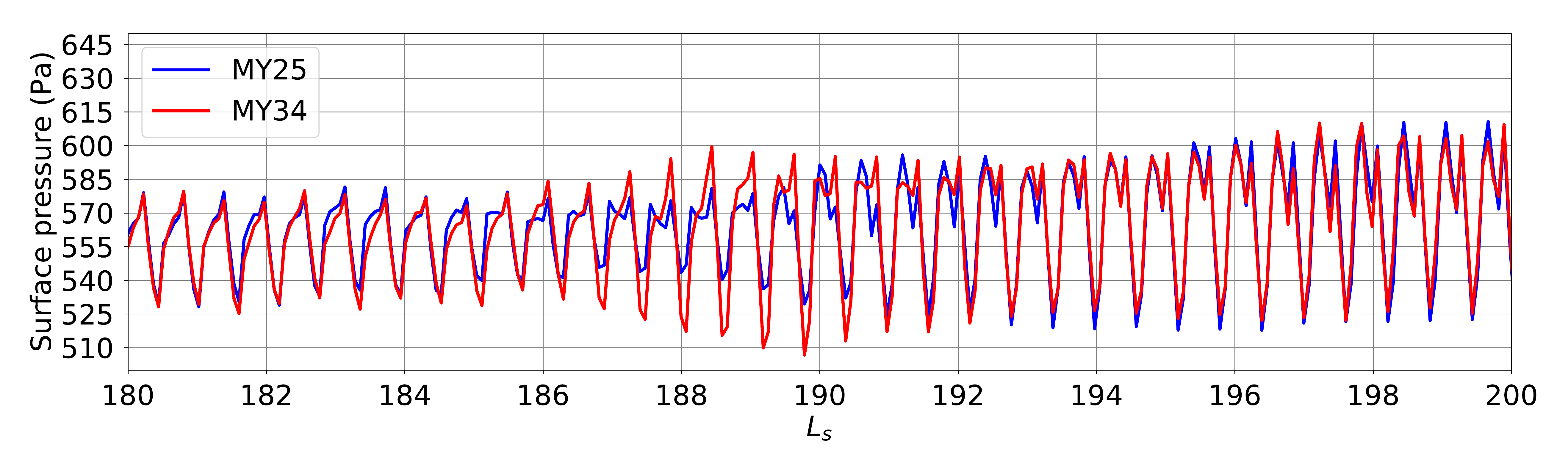}
\includegraphics[width=0.45\textwidth]{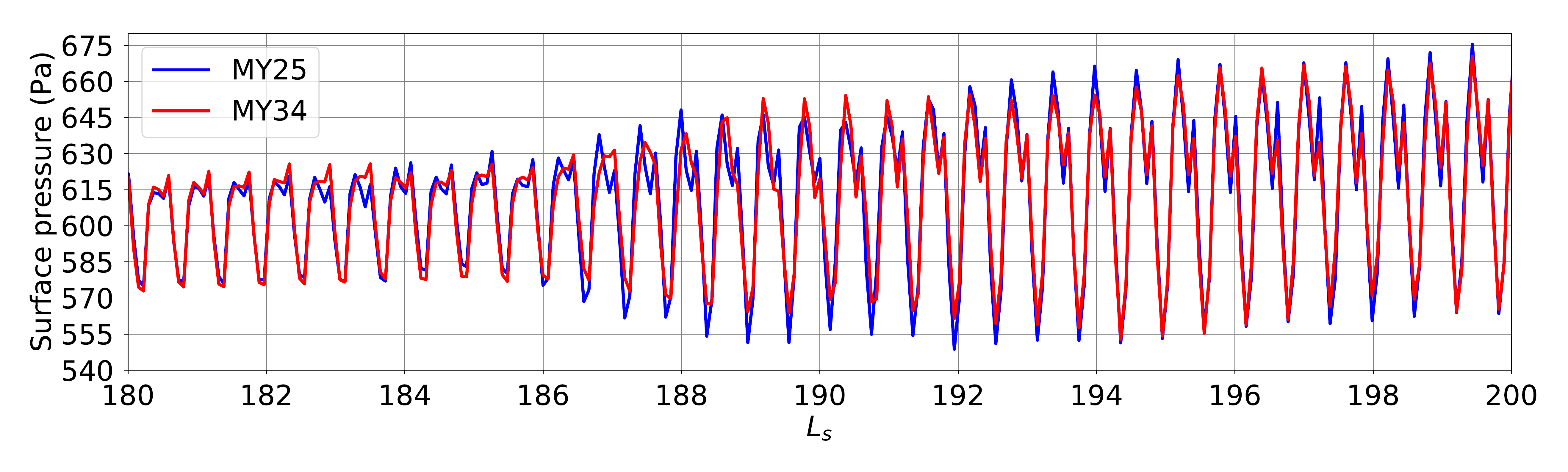}
\caption{\label{fig:daily} \emph{Daily cycle of surface pressure as simulated by the LMD Mars GCM using the MY 25 and MY 34 dust scenarios. The focus of the figure corresponds to the onset of the two GDEs. This is showing the simulated fields at the Opportunity (left) and Curiosity (right) landing sites.}}
\end{center} 
\end{figure}

\begin{figure}
    \centering
    \includegraphics[width=0.48\textwidth]{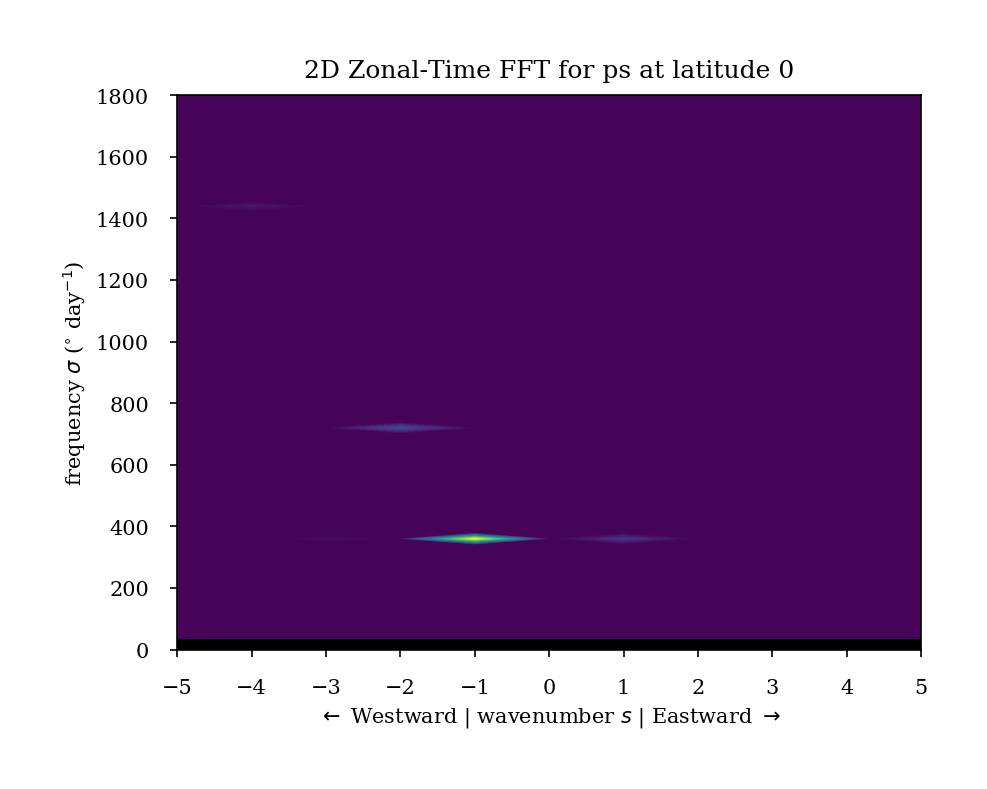}
    \includegraphics[width=0.48\textwidth]{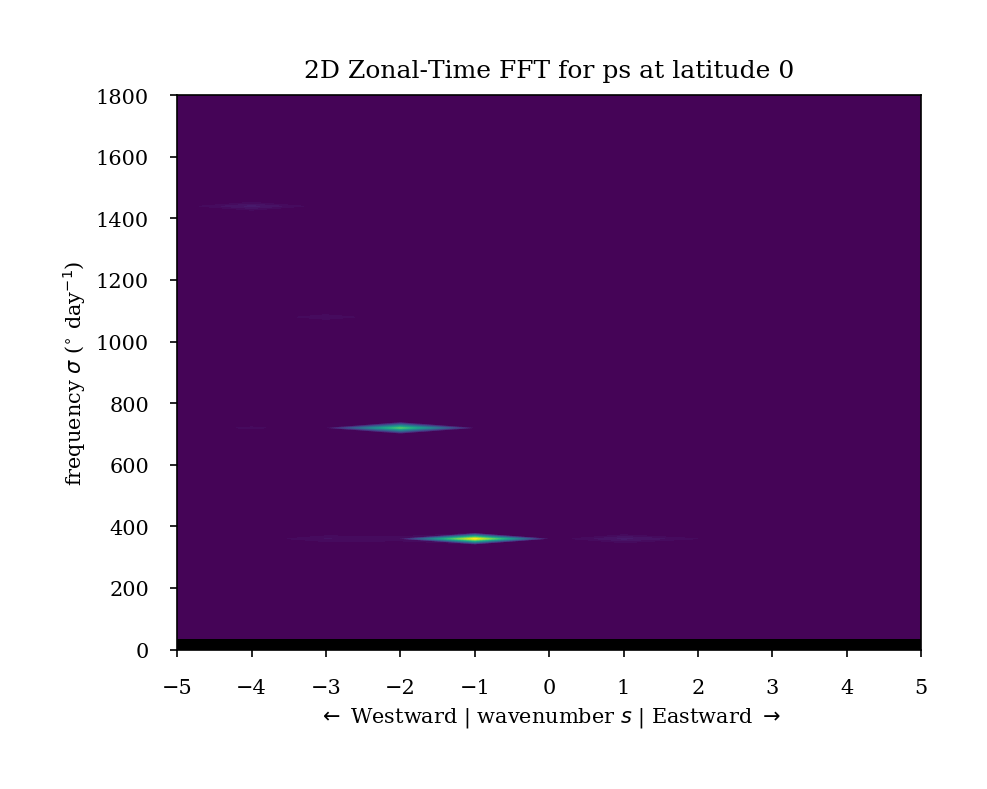}
    \caption{\label{fig:tides} \emph{Planetary waves detected by a Fast Fourier transform performed on the surface pressure simulated at the equator by the LMD Martian GCM forced by the MY 34 dust scenario. The spectral analysis is carried out on two time intervals before the MY34 GDE (sol-of-year 350 to 370, left) and during the MY 34 GDE (sol-of-year 400 to 420, right). The diurnal (-1), semi-diurnal (-2) and quarter-diurnal (-4) tides are detected, as well as the Kelvin (+1) mode. The most notable change is the reinforcement of the semi-diurnal mode by the MY 34 GDE, with an enhancement of spectral power of about one order of magnitude. During the GDE, the semi-diurnal tidal mode is as strong as the diurnal tidal mode. The quarter-diurnal mode is also significantly reinforced, becoming stronger than the wavenumber-1 Kelvin mode.}}
\end{figure}

An important test of the dynamical behavior of our LMD-MGCM simulation forced by the MY 34 dust scenario is how thermal tides react to the global increase of dust opacity following the onset of the GDE. Figure~\ref{fig:daily} shows several daily cycles of surface pressure corresponding to the onset of the MY 25 and MY 34 GDEs; Figure~\ref{fig:tides} presents a spectral analysis. Both the amplitude of the daily pressure cycle, and its morphology, are modified by the GDE at its onset. The diurnal pressure cycle is dominated by the diurnal tide before the GDE takes place. When the GDE starts to build up and the column optical depth increases, the diurnal mode increases slightly in amplitude while the semi-diurnal mode increases significantly compared to the other modes, as is already described in previous studies \citep[][their Figure 5]{Zure:93,Wils:96,Lewi:05}. The reinforcement of the semi-diurnal tide with increased column opacity is due to the fact that this tide component is dominated by a Hough mode with a large vertical wavelength \citep{Chap:70}. As a result, this Hough mode is very sensitive to forcing extended in altitude such as the absorption of incoming sunlight by dust particles during a dust storm. 

Those major changes in the tidal modes take only a couple of sols to react to the GDE onset: after their respective onsets, the GDE of MY 34 and MY 25 basically exhibit a very similar diurnal pressure cycle. This similarity between MY 34 and MY 25 is actually a good cross-calibration test for the reconstructed dust scenarios, since MY 25 is based on retrieved column optical depths from TES nadir observations, while MY 34 is based on estimated values from MCS limb observations. Interestingly, comparing the cases of MY 25 and MY 34 at the Curiosity and Opportunity landing sites in Figure~\ref{fig:daily} highlights the fact that the modification of the thermal tides is only occurring locally at the site at which the column optical depth started to rise, and not at the other site. This indicates that the extent of the storm has to become large enough for the global modification of the thermal tide to be effective.

\begin{figure}[h!] 
\begin{center}
\includegraphics[width=0.88\textwidth]{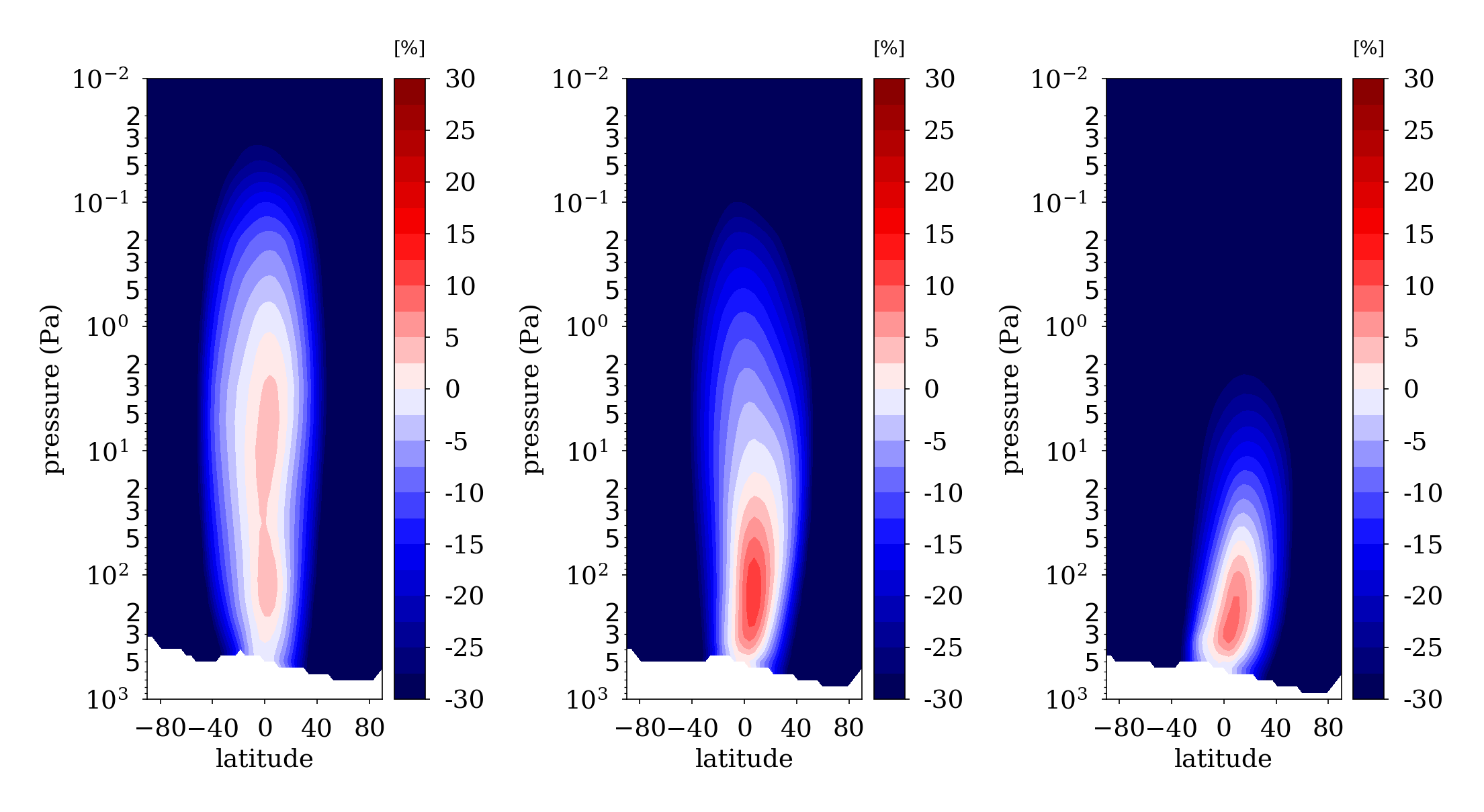}
\includegraphics[width=0.88\textwidth]{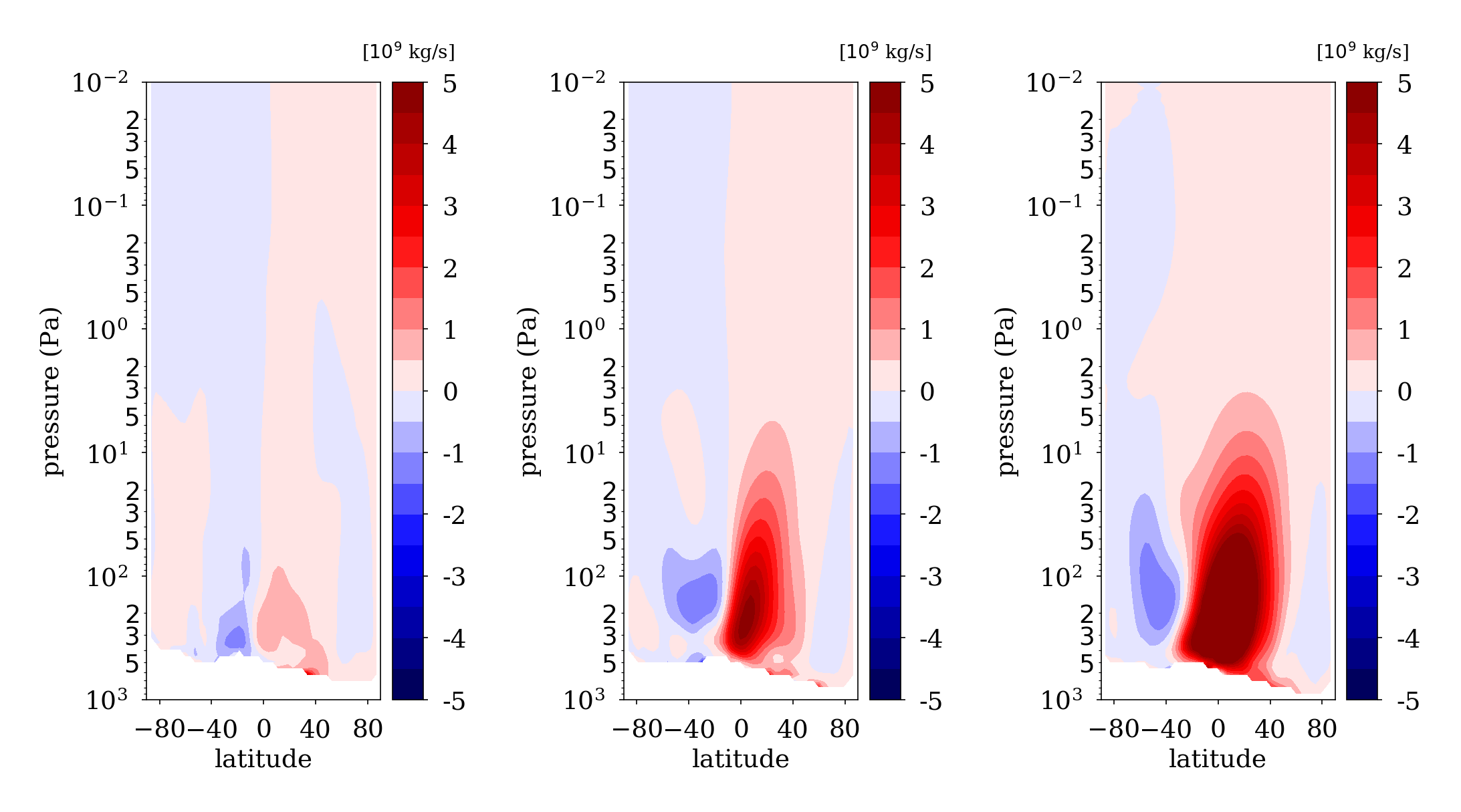}
\caption{\label{fig:superindex} \emph{The impact of the MY 34 GDE 
on the zonally-averaged global circulations on Mars 
is shown from left to right,
averaged on the $L_s$ intervals 
$[150^{\circ},180^{\circ}]$ (pre-GDE conditions), 
$[180^{\circ},210^{\circ}]$ (onset of the GDE), and
$[210^{\circ},240^{\circ}]$ (mature phase of the GDE).
[Top] Super-rotation index~$s$ computed according to \citet{Lewi:03} 
with positive values denoting regions where eastward jets are super-rotating 
i.e. exceeding the solid-body rotation of the planet.
[Bottom] Mass streamfunction with 
blue regions corresponding to counterclockwise circulation and
red regions corresponding to clockwise circulation.}}
\end{center} 
\end{figure}

The increase in column dust optical depth associated with the MY 34 GDE has a profound impact on the large-scale circulation. \citet{Lewi:03} evidenced an equatorial low-troposphere super-rotating jet in the atmosphere of Mars, and emphasized the strong positive impact of the atmospheric dust loading on this jet. Our LMD-MGCM forced simulation for MY 34 shows that the intensity of this super-rotating jet \citep[diagnosed by the super-rotating index as in][]{Lewi:03} is indeed increased following the onset of the GDE from a 5\% super-rotation index to a 15\% super-rotation index. We also find that this jet is becoming more confined close to the surface as the GDE develops (Figure~\ref{fig:superindex} top). The mean meridional circulation is also deeply impacted by the large dust loading following the onset of the MY 34 GDE: the intensity of this mean meridional circulation is enhanced by a factor of 10 following the onset and mature phase of the GDE (Figure~\ref{fig:superindex} bottom). This behaviour is similar to the evolution of the mean meridional circulation simulated under MY 25 GDE conditions \citep[e.g.,][]{Mont:05luca}.

\begin{figure}[ph!] 
\begin{center}
\includegraphics[width=0.67\textwidth]{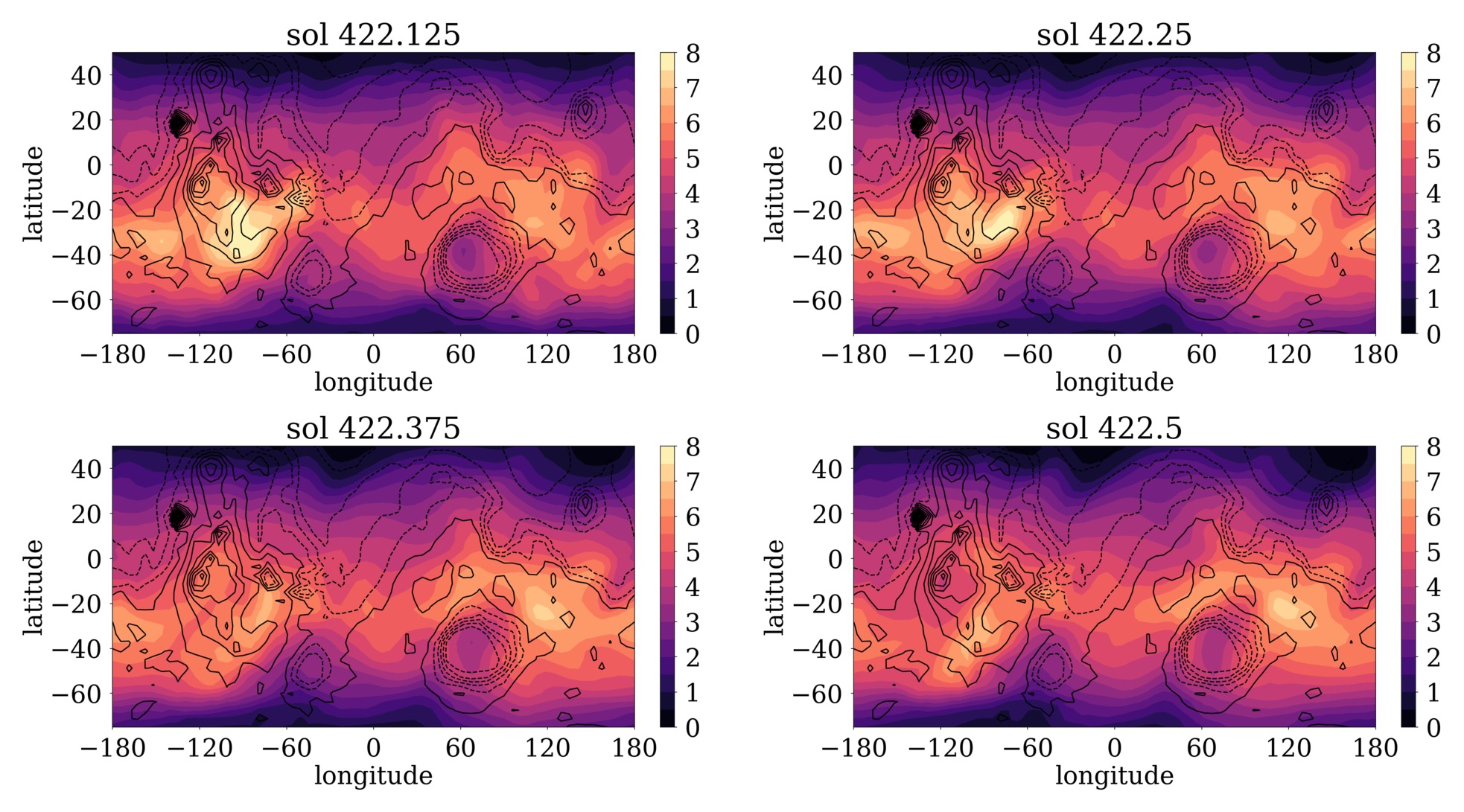}
\includegraphics[width=0.67\textwidth]{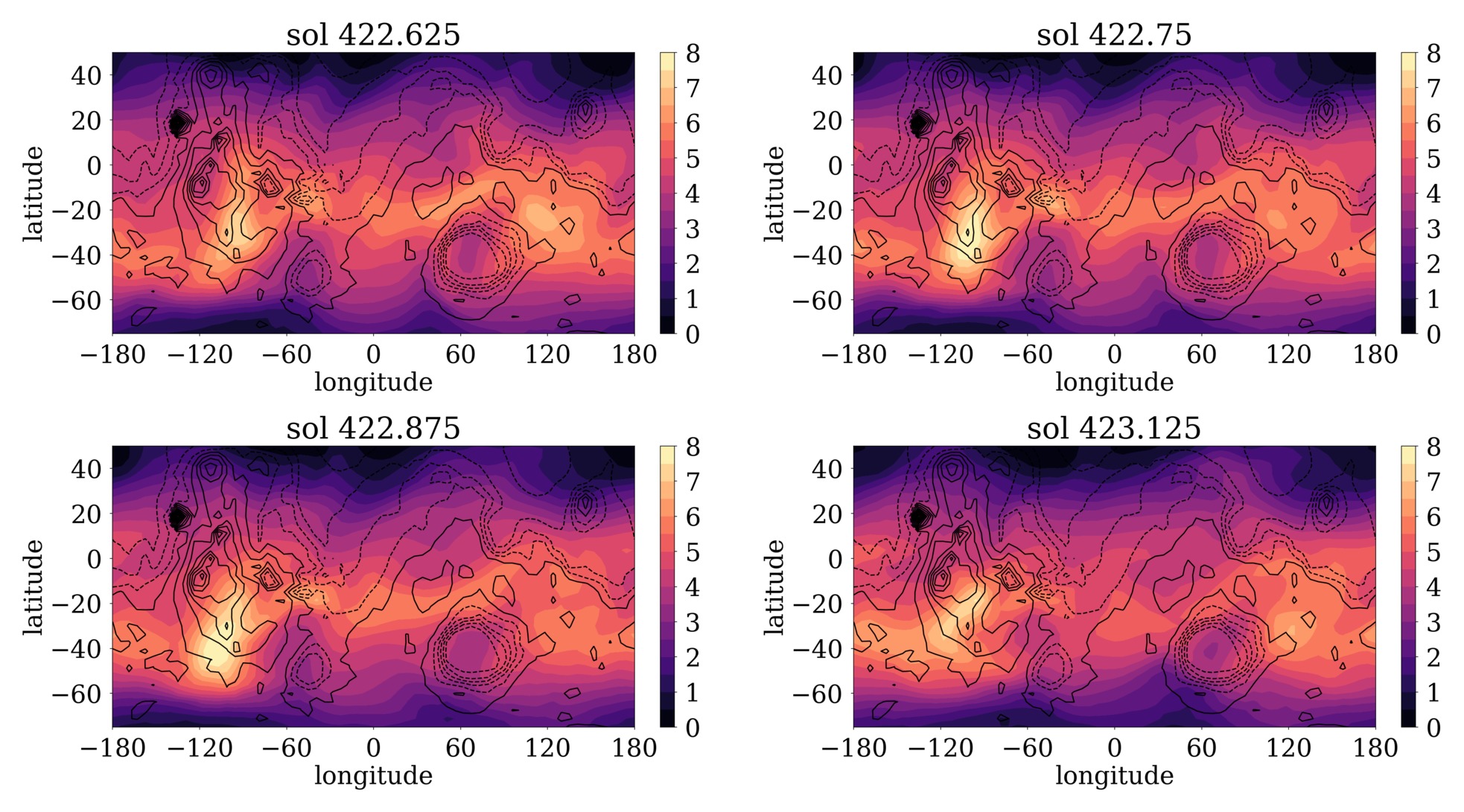}
\includegraphics[width=0.67\textwidth]{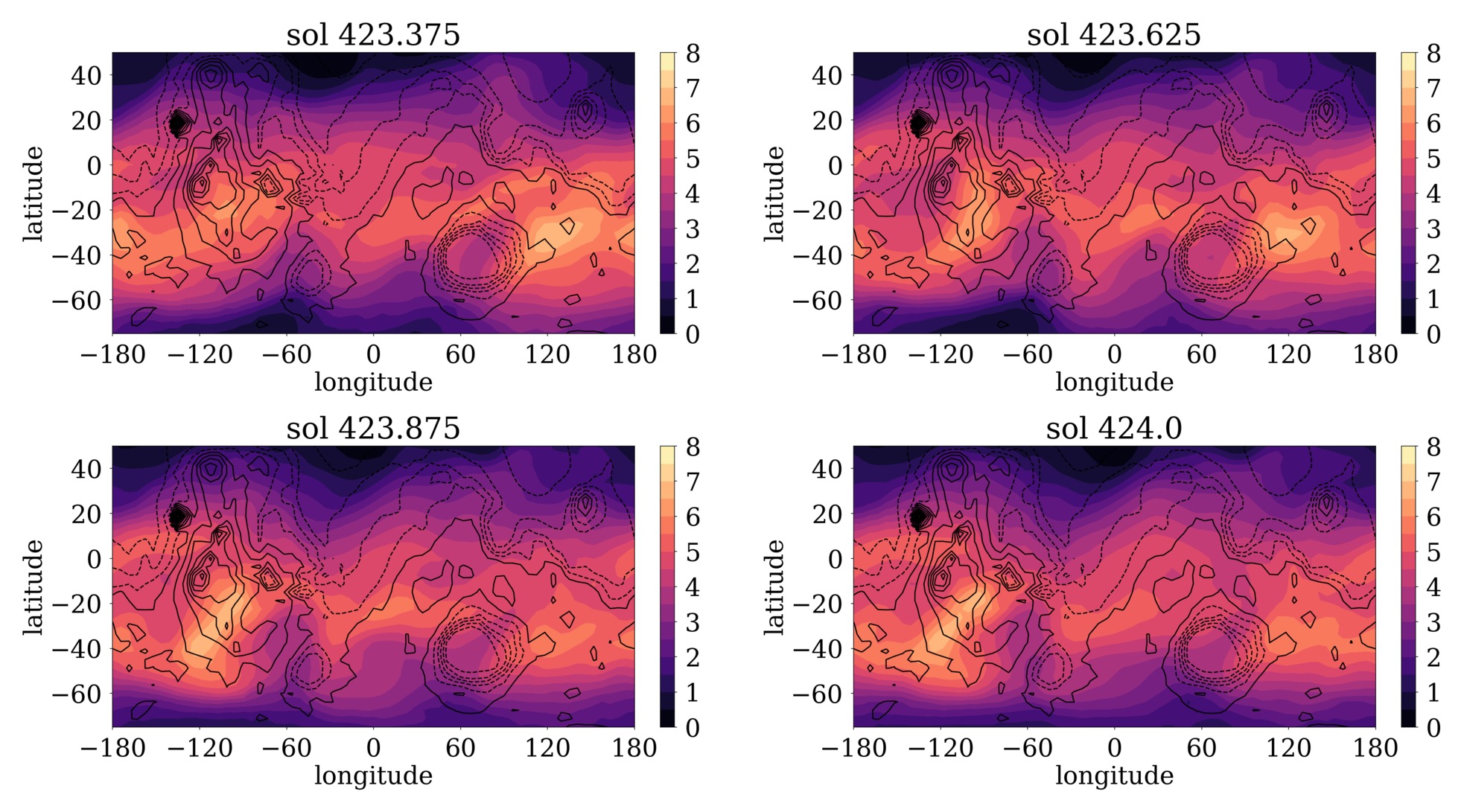}
\caption{\label{fig:diurnalvaria} \emph{Sequence of column dust optical depth maps separated by 3 hours over two sols in a LMD-MGCM ``free dust'' simulation where, contrary to the simulation forced with the MY 34 dust scenario, the dust mass mixing ratio in the model is not normalized to match the total column dust optical depth in the MY 34 dust scenario. This ``free dust'' simulation was restarted from the simulated state of the atmosphere at $L_s=210^{\circ}$ in a regular LMD-MGCM simulation forced by the MY 34 dust scenario.}}
\end{center} 
\end{figure}

Finally, we discuss the use of a ``free dust'' simulation to get some insights on the daily variability of CDOD in the GCM model. GCM simulations have already proved successful to show that large-scale circulation components (i.e. the mean meridional circulation, planetary waves, and the polar vortex) cause the vertical distribution of dust to undergo diurnal variations both in equatorial and extratropical regions [Kleinb\"ohl et al., this issue]. Figure~\ref{fig:diurnalvaria} shows the column dust optical depth simulated in a LMD-MGCM run around $L_s=210^{\circ}$ (i.e. near the peak of the MY 34 storm). The total column optical depth freely evolves in the simulation without being re-normalized using the values in the MY 34 dust scenario. As is observed in the MCS CDOD values, and by consequence in the gridded/kriged maps reconstructed following the method described in this paper, the column optical depth in the ``free dust'' model run could vary significantly on a diurnal basis in a given region. This results from horizontal transport by the large-scale circulation. While it is not possible to rule out the other possible interpretations of the observed diurnal variability of column optical depth as discussed at the end of Section~\ref{sec:diuvar}, the LMD-MGCM results in Figure~\ref{fig:diurnalvaria} strongly suggests that this variability has a physical basis, and at least part of it would be related to the large-scale horizontal transport. What our GCM simulation cannot tell is how much specific mesoscale phenomena, including dusty deep convection \citep[``rocket dust storms''][]{Spig:13rocket}, or Planetary Boundary Layer processes, would contribute to the stronger diurnal variability observed by MCS. It should also be pointed out that the GCM model uses a simplified scheme for dust lifting, which might not be well suited to reproduce strong day-night variability.

\section{Conclusions and remarks}
\label{sec:conclusions}

The work described in this paper was devoted
\begin{inparaenum}
\item to reconstruct maps of column dust optical depth for MY 34 from Mars Reconnaissance Orbiter/Mars Climate Sounder observations,
\item to analyze the seasonal, day-to-day, and daily variability of column dust showed by the maps, and
\item to use numerical simulations with the Laboratoire de M\'et\'eorologie Dynamique Mars global climate model, forced by or simply initiated with the reconstructed CDOD maps, in order to examine some aspects of the impact of the MY 34 global dust event on local and global scale dynamics, including the daily variability of column dust. 
\end{inparaenum}

The reconstructed maps for MY 34 follow the work by \citet{Mont:15} and extend the publicly available multi-annual, multi-instrument climatology of column dust optical depth to 11 Martian years. An important difference of the present work with respect to \citet{Mont:15} is that we have now reconstructed sub-daily maps of column dust, allowing us to access the analysis of the daily variability of such quantity. This was made possible by using novel retrievals (version 5.3.2) of dust opacity profiles from MCS observations during the period May 21, 2018, to October 15, 2018 ($L_s$ = [179$^\circ$, 269$^\circ$] in MY 34), which extend lower in altitude than standard version 5.2 retrievals. In general, therefore, the estimated column dust optical depth values during the global dust event of MY 34 result more accurate, within the intrinsic limitations of estimating CDODs from limb observations.

The analysis of the MY 34 column dust variability at different temporal scales using the reconstructed maps highlighted that:
\begin{itemize}
    \item MY 34 reproduced the dichotomy observed in the 10 previous years between the ``low dust loading'' and the ``high dust loading'' seasons (this might not be the case for the recently started MY 35, which featured an unusually intense dust storm during the LDL season, D. Kass, personal communication);
    \item It also featured other typical characteristics of the seasonal evolution of CDOD, such as large values at southern polar latitudes peaking at $L_s \sim 270^\circ$, a solsticial pause centered at $L_s \sim 300^\circ$, and large values peaking again at $L_s \sim 325^\circ$ during the evolution of an intense late-winter dust storm;
    \item The key distinction of MY 34 was undoubtedly the equinoctial global dust event (starting at $L_s \sim 186^\circ$ only few sols after the equivalent event in MY 25), which seemed to feature a ``storm within the storm'' at $L_s \sim 197^\circ$, boosting its growth to attain extreme characteristics, typical of GDEs;
    \item The MY 34 GDE seems also to feature very large CDOD diurnal variability at selected locations, particularly at southern mid- and high-latitudes, as already observed in the corresponding MCS dust opacity profiles by Kleinb\"ohl et al. [this issue].
\end{itemize}

While the observation of the diurnal variability in dust opacity profiles comes from direct MCS retrievals, and could be explained by global climate model simulations invoking the effects of the large-scale circulation [Kleinb\"ohl et al., this issue], the same observation in the indirectly estimated column dust values poses more challenging questions. Is the daily variability intrinsic to the column dust, or should we expect that the shape of the dust profile in the lowest one or two scale heights not directly observed by MCS (particularly in dayside observations) is not compatible with an homogeneously mixed assumption? Whether the answer leans towards the former or the latter, or a bit of both, it would open a novel view of how dust is three-dimensionally distributed within Martian dust storms.

It is not the purpose of this paper to provide a definite answer to the aforementioned question. Nevertheless, we have resorted to numerical simulations with the LMD-MGCM to provide us with some hints. Using a ``free dust'' model run initiated at $L_s$=210$^\circ$ with the reconstructed CDOD field, the model is able to reproduce some daily variability in column dust at selected locations. Despite the fact that both the range of the variability and the precise locations do not coincide with what is estimated from MCS, this result provides a physical evidence that some degree of daily variability can be expected not only in the dust profiles but also in the integrated columns. Furthermore, the dust lifting and planetary boundary layer parameterizations of the global model might actually miss some of the important features that lead to an accurate description of the three-dimensional dust distribution. The model result might, therefore, underestimate the real variability of the column dust.   

What the model simulations clearly show when forced with daily averaged CDOD maps, however, is that the impact of the MY 34 GDE on the atmospheric dynamics is as large as for the MY 25 GDE. Key features of the local and global dynamics (such as tides, mean meridional circulation, and equatorial winds) respond to the equinoctial dust events in a very similar manner. This is also an indirect validation of the MY 34 reference dust scenario based on the daily averaged CDOD maps from MCS, which is currently used in several modeling studies of the 2018 GDE. 

Future work should address the possibility of producing and making available a sub-daily dust scenario. As mentioned in Subsection~\ref{subsec:refdatabase}, we consider that this option was not currently viable, mainly because the CDOD daily variability is not yet independently confirmed, and because it is not yet clear whether model simulations forced by a sub-daily dust scenario are free of spurious effects. Current Mars GCMs might need to be adapted to handle diurnally varying CDODs in a stable and sensible fashion, if the degree of variability will be proved to be of the order of the one showed in this paper.
 
Strong emphasis should also be put in obtaining future observations of column-integrated dust as well as dust profiles with sub-daily frequency. The Planetary Fourier Spectrometer (PFS) aboard Mars Express, and the Atmospheric Chemistry Suite (ACS) aboard Trace Gas Orbiter currently provide the capability to retrieve CDOD at multiple local times, and could help in the comparison with the estimated MCS day-night variability of column dust. Future observations from the forthcoming Emirates Mars Mission (EMM) might provide even stronger evidence of the presence or absence of daily variability. 

Moreover, in order to fully characterize the diurnal cycle of dust and accurately monitor the evolution of dust storms on Mars, novel approaches must be taken in the future. These include: 
\begin{itemize}
\item the use of satellites in Mars-stationary orbits (also called ``areostationary''), which are equatorial, circular, planet-synchronous orbits equivalent to geostationary ones for the Earth \citep[see e.g.][;]{Mont:18areostat}
\item the use of instruments that allow to observe the vertical distribution of the dust in the Martian PBL, such as lidars. This is particularly important during dust storms when IR spectrometers/radiometers (both nadir- and limb-looking) fail to produce reliable retrievals because of the large atmospheric opacity and the reduced temperature contrast.
\end{itemize}


%
%
%
%
\appendix
\section{Appendix: MY 34 climatology versioning}
\label{sec:appendix}

MCS version 5.3.2 is an interim, experimental version of retrievals, leading to a possible future improved version of the whole MCS dataset. As a consequence, the MY 34 gridded and kriged datasets should be considered work in progress, as should the datasets related to other Martian years. It is our intention to regularly update the multi-annual, multi-instrument dust climatology with new observations, novel retrievals of past observations, and updated gridding methodologies/features. The updates are likely to be made publicly available on the Mars Climate Database project webpage at the URL \url{http://www-mars.lmd.jussieu.fr/}. There exist, therefore, multiple versions of these reference dust climatologies, notably for MY 34, and we would like to provide in this appendix some details about the main differences.

As mentioned in Section~\ref{sec:maps}, the reference version of the current maps from MY 24 to 32 is v2.0. For this version, the used gridding/kriging methodology is precisely the one described in \citet{Mont:15}. Version 2.1 is a specific version only for MY 33, where we have used a different weight for THEMIS observations with respect to MCS observations, in order to account for THEMIS retrievals provided at later local times, and we started using MCS v5.2 ``two-dimensional'' retrievals \citep{Klei:17} instead of v4.3 ``one-dimensional'' retrievals for previous years 28 to 32. 

For MY 34, we have produced three intermediate versions (v2.2, v2.3, and v2.4) and the v2.5 described in this paper, which should be considered as the reference version. All three intermediate versions use MCS v5.3.2 retrievals for the available period, and MCS v5.2 for the rest of the time, but do not use the two distinctive features described in Subsection~\ref{subsec:gridmethod}, namely the local time cut-off window of $\pm7$~hours for observations considered for the weighted average at each grid point, and the 6~hour moving average producing 4 maps per sol. Instead, they use observations at all local times for each grid point (except in v2.3 and v2.4 during the GDE, see below), and the 24~h moving average produces only one map per sol, centered at MUT=12:00, as described in \citet{Mont:15}. 

Within the intermediate versions, the differences are as following:
\begin{itemize}
    \item v2.2: This version still uses the same data QC and gridding methodology as in \citet{Mont:15}. The use of dayside values is limited by the application of the ``dayside'' filter with 8 km cut-off altitude threshold at any time. Apart the use of MCS v5.3.2 retrievals, the only other difference with respect to v2.0 is in the kriged maps, where we have artificially introduced climatological south cap edge "storms" only for solar longitude earlier than 180$^\circ$, as in MY 25. This version also uses MCS observations only until end of September 2018 ($L_s \sim 260^\circ$), stopping at SOY 501.
    \item v2.3: This version uses only dayside values during the GDE ($186.5^\circ < L_s < 269^\circ$, SOY 383 to 515). It has also an improved data QC with respect to v2.2: we introduced the ``water ice'' filter, the ``cross-track'' filter, and we did not apply the ``dayside filter'' with 8 km cut-off altitude threshold during the GDE and the late winter regional storm ($L_s > 312^\circ$). This allowed to use many more dayside values during the two major dust events of MY 34, increasing the overall optical depth to levels observed by, e.g., Opportunity rover. We also redefined the estimation of uncertainties according to the scheme that was later adopted in v2.5 (but with slightly lower uncertainties overall). Furthermore, we changed a couple of parameters in the IWB methodology: the criterion to accept a value of weighted average at a particular grid point at any given iteration became that there must be at least one observation within a distance of 200 km from the grid point. We started using the same surface pressure recorded in the MCS dataset to renormalize CDOD to 610 Pa, instead of the MCD surface pressure. If MCS surface pressure is not retrieved, we associated a 10\% uncertainty by default. We stopped using the artificial modification of a latitude band around the southern polar cap at all times. This version also uses MCS observations only until end of February 2018 ($L_s \sim 349^\circ$), stopping at SOY 647.
    \item v2.4: This version is quite similar to v2.3. The only differences are in the refined data QC, which is the one we also use in v2.5 (see Subsection~\ref{subsec:QC}). It also extends until the end of MY 34.   
\end{itemize}

\begin{figure}[ph!] 
\begin{center}
\includegraphics[width=1.\textwidth]{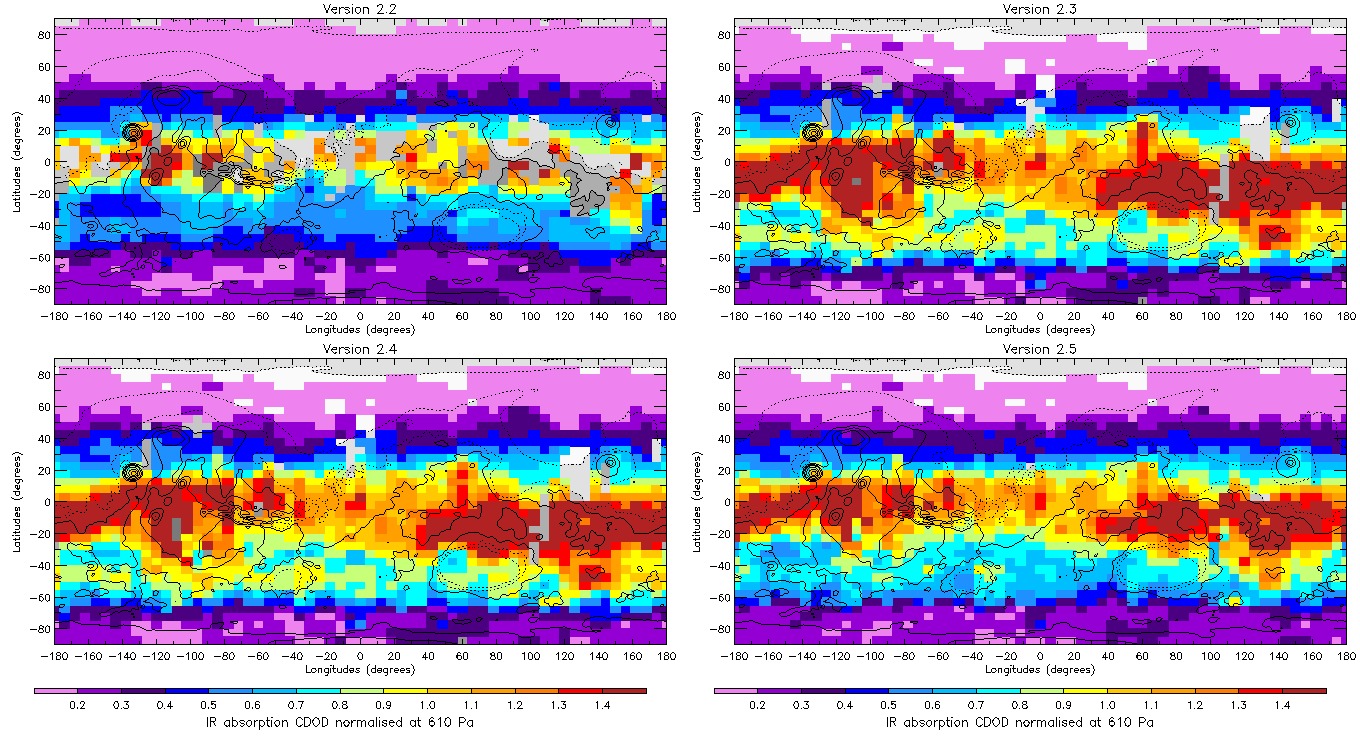}
\caption{\label{fig:gridcomparvers} \emph{Different versions of the gridded maps of 9.3~$\mu$m absorption column dust optical depth for SOY 400, $L_s \sim 196^{\circ}$, in the growing phase of the Global Dust Event of MY 34. V2.2 is the top left map, v2.3 is the top right one, v2.4 is the bottom left one, and the reference v2.5 is the bottom right one. The spatial resolution of all gridded maps is 6$^\circ$ longitude $\times$ 5$^\circ$ latitude. }}
\end{center} 
\end{figure}

\begin{figure}[ht] 
\begin{center}
\includegraphics[width=1.\textwidth]{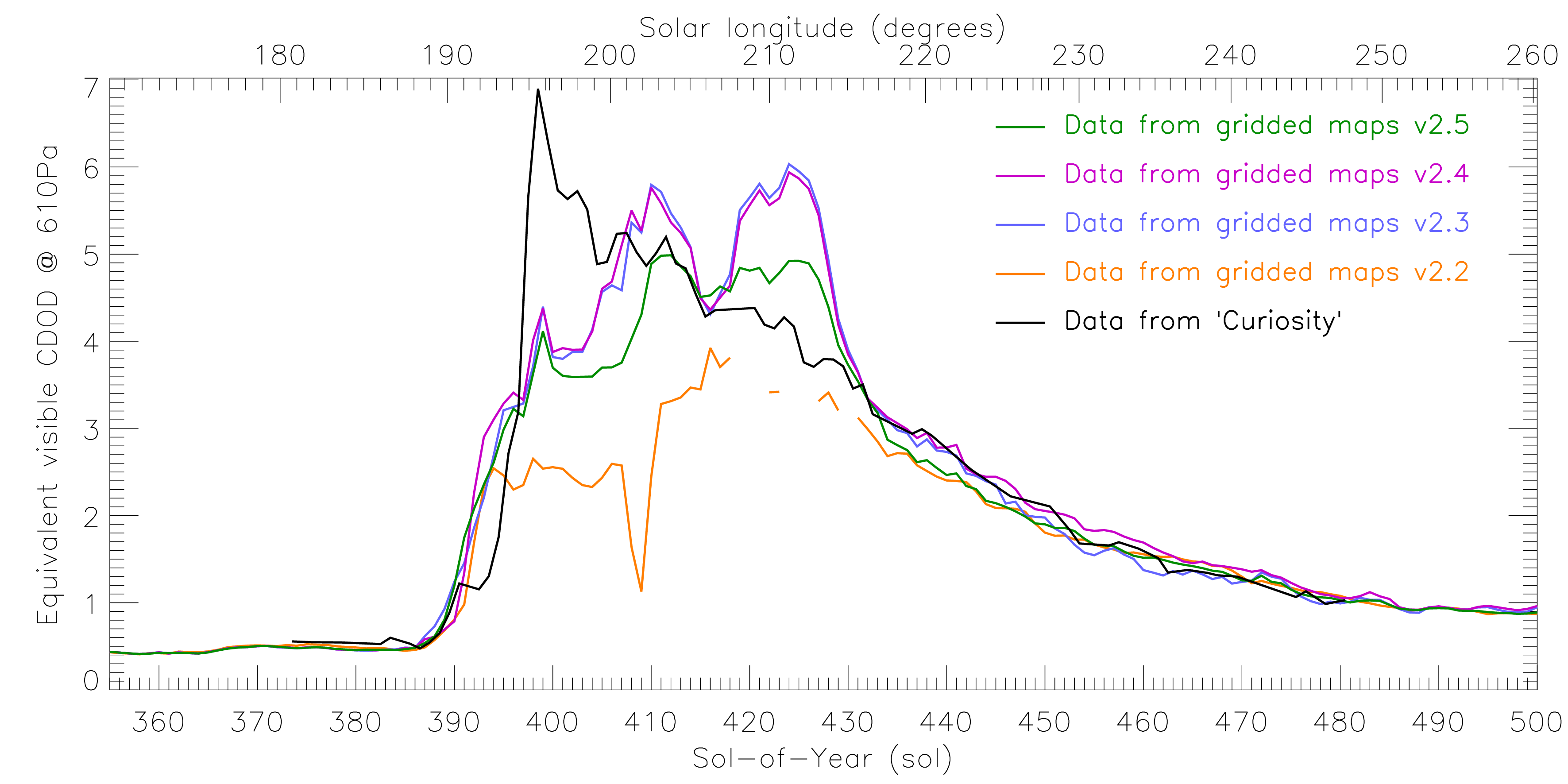}
\caption{\label{fig:comparverscuriosity} \emph{Time series of equivalent visible column dust optical depth calculated from the 9.3~$\mu$m absorption CDOD normalized to 610 Pa, extracted from different versions of the  gridded maps in an area around Gale crater, compared to the time series of visible column optical depth measured by MastCAM aboard NASA's ``Curiosity'' rover (black line). Curiosity observations [Guzevich et al., 2018, see main text] have been daily averaged and normalized to 610 Pa (using the surface pressure from the Mars Climate Database \texttt{pres0} routine). Both time series are shown between Sol-of-Year 355 and 500, i.e. $L_s \sim [170^{\circ}, 260^{\circ}]$. We used a factor of 2.6 to convert 9.3~$\mu m$ absorption CDOD into equivalent visible ones. Data from gridded maps are averaged in longitude=[123$^{\circ}$E, 153$^{\circ}$E], latitude=[15$^{\circ}$S, 10$^{\circ}$N] centered around Curiosity landing site at longitude $137.4^{\circ}$E and latitude $4.6^{\circ}$S.}}
\end{center} 
\end{figure}

Refer to Figures~\ref{fig:gridcomparvers} and \ref{fig:comparverscuriosity} for a comparison of results using versions 2.2, 2.3, 2.4 and 2.5.
 
%
%

%

%

\section*{Acknowledgments}
\label{sec:ack}

The work related to the production of reconstructed gridded and kriged CDOD maps for MY 34 is funded by the French Centre National d'Etudes Spatiales (CNES). This work uses technical achievements and expertise obtained during a parallel project of improving the gridding of column dust optical depth retrievals from satellite observations, funded by NASA PDART programme (Grant no. NNX15AN06G). 

The maps of gridded and kriged CDOD produced in this work will be publicly available under the Mars Climate Database project (at the URL \url{http://www-mars.lmd.jussieu.fr/}) after the version of this manuscript submitted to the Journal of Geophysical Research - Planets is accepted for publication. A beta version of these maps is currently available upon request to the corresponding author (lmontabone@spacescience.org). 
Mars Climate Sounder data is publicly available on NASA's Planetary Data System (https://pds-atmospheres.nmsu.edu/). Work related to MCS observations and retrievals (including the estimates of CDOD) is carried out at the Jet Propulsion Laboratory, California Institute of Technology, and is performed under a contract with NASA. Government sponsorship is acknowledged.
Data from Curiosity rover obtained by the Rover Environmental Monitoring Station (REMS) instrument is publicly available as supplementary material of \citet{Guze:18}.

The authors wish to thank R. John Wilson for useful comments on early versions of the reference MY 34 dust climatology datasets. 

Luca Montabone wishes to dedicate this work to his father Augusto, who started his journey to the stars during the production of this paper.


\begin{thebibliography}{42}
\providecommand{\natexlab}[1]{#1}
\providecommand{\url}[1]{\texttt{#1}}
\expandafter\ifx\csname urlstyle\endcsname\relax
  \providecommand{\doi}[1]{doi: #1}\else
  \providecommand{\doi}{doi: \begingroup \urlstyle{rm}\Url}\fi

\bibitem[{Cantor}(2007)]{Cant:07}
B.~A. {Cantor}.
\newblock {MOC observations of the 2001 Mars planet-encircling dust storm}.
\newblock \emph{Icarus}, 186:\penalty0 60--96, 2007.
\newblock \doi{10.1016/j.icarus.2006.08.019}.

\bibitem[{Chapman} and {Lindzen}(1970)]{Chap:70}
S.~{Chapman} and R.~{Lindzen}.
\newblock \emph{{Atmospheric tides. Thermal and gravitational}}.
\newblock Dordrecht: Reidel, 1970, 1970.

\bibitem[{Cola{\"i}tis} et~al.(2013){Cola{\"i}tis}, {Spiga}, {Hourdin}, {Rio},
  {Forget}, and {Millour}]{Cola:13}
A.~{Cola{\"i}tis}, A.~{Spiga}, F.~{Hourdin}, C.~{Rio}, F.~{Forget}, and
  E.~{Millour}.
\newblock {A thermal plume model for the Martian convective boundary layer}.
\newblock \emph{Journal of Geophysical Research (Planets)}, 118:\penalty0
  1468--1487, 2013.
\newblock \doi{10.1002/jgre.20104}.

\bibitem[{Fedorova} et~al.(2018){Fedorova}, {Bertaux}, {Betsis}, {Montmessin},
  {Korablev}, {Maltagliati}, and {Clarke}]{Fedo:18}
A.~{Fedorova}, J.-L. {Bertaux}, D.~{Betsis}, F.~{Montmessin}, O.~{Korablev},
  L.~{Maltagliati}, and J.~{Clarke}.
\newblock {Water vapor in the middle atmosphere of Mars during the 2007 global
  dust storm}.
\newblock \emph{Icarus}, 300:\penalty0 440--457, 2018.
\newblock \doi{10.1016/j.icarus.2017.09.025}.

\bibitem[{Forget} et~al.(2007){Forget}, {Spiga}, {Dolla}, {Vinatier},
  {Melchiorri}, {Drossart}, {Gendrin}, {Bibring}, {Langevin}, and
  {Gondet}]{Forg:07omeg}
F.~{Forget}, A.~{Spiga}, B.~{Dolla}, S.~{Vinatier}, R.~{Melchiorri},
  P.~{Drossart}, A.~{Gendrin}, J.-P. {Bibring}, Y.~{Langevin}, and B.~{Gondet}.
\newblock {Remote sensing of surface pressure on Mars with the Mars
  Express/OMEGA spectrometer: 1. Retrieval method}.
\newblock \emph{Journal of Geophysical Research (Planets)}, 112\penalty0
  (E11):\penalty0 8--+, 2007.
\newblock \doi{10.1029/2006JE002871}.

\bibitem[{Gonz{\'a}lez-Galindo} et~al.(2011){Gonz{\'a}lez-Galindo},
  {M{\"a}{\"a}tt{\"a}nen}, {Forget}, and {Spiga}]{Gonz:11}
F.~{Gonz{\'a}lez-Galindo}, A.~{M{\"a}{\"a}tt{\"a}nen}, F.~{Forget}, and
  A.~{Spiga}.
\newblock The martian mesosphere as revealed by co2 clouds observations and
  general circulation modeling.
\newblock \emph{Icarus}, 216:\penalty0 10--22, 2011.

\bibitem[Guzewich et~al.(2018)Guzewich, Lemmon, Smith, Mart{\'i}nez,
  de~Vicente-Retortillo, Newman, Baker, Campbell, Cooper, G{\'o}mez-Elvira,
  Harri, Hassler, Martin-Torres, McConnochie, Moores, Kahanp{\"a}{\"a}, Khayat,
  Richardson, Smith, Sullivan, de~la Torre~Juarez, Vasavada,
  Vi{\'u}dez-Moreiras, Zeitlin, and Mier]{Guze:18}
Scott~D. Guzewich, M.~Lemmon, C.~L. Smith, G.~Mart{\'i}nez,
  A.~de~Vicente-Retortillo, C.~E. Newman, M.~Baker, C.~Campbell, B.~Cooper,
  J.~G{\'o}mez-Elvira, A.-M. Harri, D.~Hassler, F.~J. Martin-Torres,
  T.~McConnochie, J.~E. Moores, H.~Kahanp{\"a}{\"a}, A.~Khayat, M.~I.
  Richardson, M.~D. Smith, R.~Sullivan, M.~de~la Torre~Juarez, A.~R. Vasavada,
  D.~Vi{\'u}dez-Moreiras, C.~Zeitlin, and Maria-Paz~Zorzano Mier.
\newblock Mars science laboratory observations of the 2018/mars year 34 global
  dust storm.
\newblock \emph{Geophysical Research Letters}, 0\penalty0 (ja), 2018.
\newblock \doi{10.1029/2018GL080839}.
\newblock URL
  \url{https://agupubs.onlinelibrary.wiley.com/doi/abs/10.1029/2018GL080839}.

\bibitem[{Heavens} et~al.(2018){Heavens}, {Kleinb{\"o}hl}, {Chaffin},
  {Halekas}, {Kass}, {Hayne}, {McCleese}, {Piqueux}, {Shirley}, and
  {Schofield}]{Heav:18}
N.~G. {Heavens}, A.~{Kleinb{\"o}hl}, M.~S. {Chaffin}, J.~S. {Halekas}, D.~M.
  {Kass}, P.~O. {Hayne}, D.~J. {McCleese}, S.~{Piqueux}, J.~H. {Shirley}, and
  J.~T. {Schofield}.
\newblock {Hydrogen escape from Mars enhanced by deep convection in dust
  storms}.
\newblock \emph{Nature Astronomy}, 2:\penalty0 126--132, 2018.
\newblock \doi{10.1038/s41550-017-0353-4}.

\bibitem[{Kahre} et~al.(2017){Kahre}, {Murphy}, {Newman}, {Wilson}, {Cantor},
  {Lemmon}, and {Wolff}]{Kahr:17}
Melinda~A. {Kahre}, James~R. {Murphy}, Claire~E. {Newman}, R.~John {Wilson},
  Bruce~A. {Cantor}, Mark~T. {Lemmon}, and Michael~J. {Wolff}.
\newblock \emph{{The Mars Dust Cycle}}, pages 229--294.
\newblock 2017.
\newblock \doi{10.1017/9781139060172.010}.

\bibitem[{Kass} et~al.(2016){Kass}, {Kleinb{\"o}hl}, {McCleese}, {Schofield},
  and {Smith}]{Kass:16}
D.~M. {Kass}, A.~{Kleinb{\"o}hl}, D.~J. {McCleese}, J.~T. {Schofield}, and
  M.~D. {Smith}.
\newblock {Interannual similarity in the Martian atmosphere during the dust
  storm season}.
\newblock \emph{Geophysical Research Letters}, 43\penalty0 (12):\penalty0
  6111--6118, Jun 2016.
\newblock \doi{10.1002/2016GL068978}.

\bibitem[{Kleinb{\"o}hl} et~al.(2009){Kleinb{\"o}hl}, {Schofield}, {Kass},
  {Abdou}, {Backus}, {Sen}, {Shirley}, {Lawson}, {Richardson}, {Taylor},
  {Teanby}, and {McCleese}]{Klei:09}
A.~{Kleinb{\"o}hl}, J.~T. {Schofield}, D.~M. {Kass}, W.~A. {Abdou}, C.~R.
  {Backus}, B.~{Sen}, J.~H. {Shirley}, W.~G. {Lawson}, M.~I. {Richardson},
  F.~W. {Taylor}, N.~A. {Teanby}, and D.~J. {McCleese}.
\newblock {Mars Climate Sounder limb profile retrieval of atmospheric
  temperature, pressure, and dust and water ice opacity}.
\newblock \emph{Journal of Geophysical Research (Planets)}, 114:\penalty0
  E10006, October 2009.
\newblock \doi{10.1029/2009JE003358}.

\bibitem[Kleinb{\"o}hl et~al.(2011)Kleinb{\"o}hl, Schofield, Abdou, Irwin, and
  de~Kok]{Klei:11}
A.~Kleinb{\"o}hl, J.~T. Schofield, W.~A. Abdou, P.~G.~J. Irwin, and R.~J.
  de~Kok.
\newblock A single-scattering approximation for infrared radiative transfer in
  limb geometry in the {M}artian atmosphere.
\newblock \emph{J. Quant. Spectrosc. Rad. Transfer}, 112:\penalty0 1568--1580,
  2011.

\bibitem[{Kleinb{\"o}hl} et~al.(2017b){Kleinb{\"o}hl}, {Chen}, and
  {Schofield}]{Klei:17mamo}
A.~{Kleinb{\"o}hl}, L.~{Chen}, and J.~T. {Schofield}.
\newblock {Far Infrared Spectroscopic Parameters of Mars Atmospheric Aerosols
  and their Application to MCS Retrievals in High Aerosol Conditions}.
\newblock In F.~{Forget} and M.~{Millour}, editors, \emph{The Mars Atmosphere:
  Modelling and observation}, page 2230, January 2017b.

\bibitem[{Kleinb{\"o}hl} et~al.(2013){Kleinb{\"o}hl}, {John Wilson}, {Kass},
  {Schofield}, and {McCleese}]{Klei:13}
Armin {Kleinb{\"o}hl}, R.~{John Wilson}, David {Kass}, John~T. {Schofield}, and
  Daniel~J. {McCleese}.
\newblock {The semidiurnal tide in the middle atmosphere of Mars}.
\newblock \emph{Geophysical Research Letters}, 40\penalty0 (10):\penalty0
  1952--1959, May 2013.
\newblock \doi{10.1002/grl.50497}.

\bibitem[Kleinböhl et~al.(2017a)Kleinböhl, Friedson, and Schofield]{Klei:17}
A.~Kleinböhl, A.~J. Friedson, and J.~T. Schofield.
\newblock Two-dimensional radiative transfer for the retrieval of limb emission
  measurements in the martian atmosphere.
\newblock \emph{Journal of Quantitative Spectroscopy and Radiative Transfer},
  187:\penalty0 511 -- 522, 2017a.
\newblock ISSN 0022-4073.
\newblock \doi{https://doi.org/10.1016/j.jqsrt.2016.07.009}.
\newblock URL
  \url{http://www.sciencedirect.com/science/article/pii/S0022407316302667}.

\bibitem[{Levine} et~al.(2018){Levine}, {Winterhalter}, and
  {Kerschmann}]{Levi:18book}
J.~S. {Levine}, D.~{Winterhalter}, and R.~L. {Kerschmann}.
\newblock \emph{Dust in the Atmosphere of Mars and its Impact on Human
  Exploration}.
\newblock Cambridge Scholars Publishing, UK, 2018.

\bibitem[{Lewis} and {Barker}(2005)]{Lewi:05}
S.~R. {Lewis} and P.~R. {Barker}.
\newblock {Atmospheric tides in a Mars general circulation model with data
  assimilation}.
\newblock \emph{Advances in Space Research}, 36:\penalty0 2162--2168, 2005.
\newblock \doi{10.1016/j.asr.2005.05.122}.

\bibitem[{Lewis} and {Read}(2003)]{Lewi:03}
S.~R. {Lewis} and P.~L. {Read}.
\newblock {Equatorial jets in the dusty Martian atmosphere}.
\newblock \emph{Journal of Geophysical Research (Planets)}, 108:\penalty0
  5034--+, 2003.
\newblock \doi{10.1029/2002JE001933}.

\bibitem[{Lewis} et~al.(2016){Lewis}, {Mulholland}, {Read}, {Montabone},
  {Wilson}, and {Smith}]{Lewi:16}
S.~R. {Lewis}, D.~P. {Mulholland}, P.~L. {Read}, L.~{Montabone}, R.~J.
  {Wilson}, and M.~D. {Smith}.
\newblock {The solsticial pause on Mars: 1. A planetary wave reanalysis}.
\newblock \emph{Icarus}, 264:\penalty0 456--464, 2016.
\newblock \doi{10.1016/j.icarus.2015.08.039}.

\bibitem[{Madeleine} et~al.(2011){Madeleine}, {Forget}, {Millour}, {Montabone},
  and {Wolff}]{Made:11}
J.-B. {Madeleine}, F.~{Forget}, E.~{Millour}, L.~{Montabone}, and M.~J.
  {Wolff}.
\newblock {Revisiting the radiative impact of dust on Mars using the LMD Global
  Climate Model}.
\newblock \emph{Journal of Geophysical Research (Planets)}, 116:\penalty0
  E11010, November 2011.
\newblock \doi{10.1029/2011JE003855}.

\bibitem[{Madeleine} et~al.(2012){Madeleine}, {Forget}, {Millour}, {Navarro},
  and {Spiga}]{Made:12radclouds}
J.-B. {Madeleine}, F.~{Forget}, E.~{Millour}, T.~{Navarro}, and A.~{Spiga}.
\newblock {The influence of radiatively active water ice clouds on the Martian
  climate}.
\newblock \emph{Geophys.~Res.~Lett.}, 39:\penalty0 L23202, 2012.
\newblock \doi{10.1029/2012GL053564}.

\bibitem[Malin et~al.(2018)Malin, Cantor, and W.]{Mali:18rds}
M.~C. Malin, B.~A. Cantor, and Britton~A. W.
\newblock Mro marci weather report for the week of 4 june 2018 – 10 june
  2018.
\newblock Malin Space Science Systems Captioned Image Release, MSSS-534,
  \url{http://www.msss.com/msss_images/2018/06/13/}, 2018.

\bibitem[Martin and Zurek(1993)]{Mart:93leo}
Leonard~J. Martin and Richard~W. Zurek.
\newblock An analysis of the history of dust activity on {Mars}.
\newblock \emph{J.~Geophys.~Res.}, 98\penalty0 (E2):\penalty0 3221--3246, 1993.

\bibitem[{McCleese} et~al.(2007){McCleese}, {Schofield}, {Taylor}, {Calcutt},
  {Foote}, {Kass}, {Leovy}, {Paige}, {Read}, and {Zurek}]{Mccl:07}
D.~J. {McCleese}, J.~T. {Schofield}, F.~W. {Taylor}, S.~B. {Calcutt}, M.~C.
  {Foote}, D.~M. {Kass}, C.~B. {Leovy}, D.~A. {Paige}, P.~L. {Read}, and R.~W.
  {Zurek}.
\newblock {Mars Climate Sounder: An investigation of thermal and water vapor
  structure, dust and condensate distributions in the atmosphere, and energy
  balance of the polar regions}.
\newblock \emph{Journal of Geophysical Research (Planets)}, 112\penalty0
  (E11):\penalty0 5--+, 2007.
\newblock \doi{10.1029/2006JE002790}.

\bibitem[{Millour} et~al.(2015){Millour}, {Forget}, {Spiga}, {Navarro},
  {Madeleine}, {Montabone}, {Pottier}, {Lefevre}, {Montmessin}, {Chaufray},
  {Lopez-Valverde}, {Gonzalez-Galindo}, {Lewis}, {Read}, {Huot}, {Desjean}, and
  {MCD/GCM development Team}]{Mill:15}
E.~{Millour}, F.~{Forget}, A.~{Spiga}, T.~{Navarro}, J.-B. {Madeleine},
  L.~{Montabone}, A.~{Pottier}, F.~{Lefevre}, F.~{Montmessin}, J.-Y.
  {Chaufray}, M.~A. {Lopez-Valverde}, F.~{Gonzalez-Galindo}, S.~R. {Lewis},
  P.~L. {Read}, J.-P. {Huot}, M.-C. {Desjean}, and {MCD/GCM development Team}.
\newblock {The Mars Climate Database (MCD version 5.2)}.
\newblock \emph{European Planetary Science Congress 2015}, 10:\penalty0
  EPSC2015--438, 2015.

\bibitem[{Montabone} and {Forget}(2018)]{Mont:18book}
L.~{Montabone} and F.~{Forget}.
\newblock \emph{{Forecasting Dust Storms on Mars: a Short Review}}, page 000.
\newblock Cambridge Scholars Publishing, Ed. {Levine}, G.~S. and
  {Winterhalter}, D. and {Kerschmann}, R.~L., 2018.
\newblock \doi{doi_tbd}.

\bibitem[{Montabone} et~al.(2005){Montabone}, {Lewis}, and {Read}]{Mont:05luca}
L.~{Montabone}, S.~R. {Lewis}, and P.~L. {Read}.
\newblock {Interannual variability of Martian dust storms in assimilation of
  several years of Mars global surveyor observations}.
\newblock \emph{Advances in Space Research}, 36:\penalty0 2146--2155, 2005.
\newblock \doi{10.1016/j.asr.2005.07.047}.

\bibitem[{Montabone} et~al.(2015){Montabone}, {Forget}, {Millour}, {Wilson},
  {Lewis}, {Cantor}, {Kass}, {Kleinb{\"o}hl}, {Lemmon}, {Smith}, and
  {Wolff}]{Mont:15}
L.~{Montabone}, F.~{Forget}, E.~{Millour}, R.~J. {Wilson}, S.~R. {Lewis},
  B.~{Cantor}, D.~{Kass}, A.~{Kleinb{\"o}hl}, M.~T. {Lemmon}, M.~D. {Smith},
  and M.~J. {Wolff}.
\newblock {Eight-year climatology of dust optical depth on Mars}.
\newblock \emph{Icarus}, 251:\penalty0 65--95, 2015.
\newblock \doi{10.1016/j.icarus.2014.12.034}.

\bibitem[{Montabone} et~al.(2018){Montabone}, {Forget}, {Smith}, {Cantor},
  {Wolff}, {Capderou}, and {VanWoerkom}]{Mont:18areostat}
Luca {Montabone}, Francois {Forget}, Michael {Smith}, Bruce {Cantor}, Michael
  {Wolff}, Michel {Capderou}, and Michael {VanWoerkom}.
\newblock {Mars Aerosol Tracker (MAT): An Areostationary SmallSat to Monitor
  Dust Storms and Water Ice Clouds}.
\newblock In \emph{42nd COSPAR Scientific Assembly}, volume~42, pages
  B0.2--16--18, Jul 2018.

\bibitem[{Mulholland} et~al.(2013){Mulholland}, {Read}, and {Lewis}]{Mulh:13}
D.~P. {Mulholland}, P.~L. {Read}, and S.~R. {Lewis}.
\newblock {Simulating the interannual variability of major dust storms on Mars
  using variable lifting thresholds}.
\newblock \emph{Icarus}, 223:\penalty0 344--358, 2013.
\newblock \doi{10.1016/j.icarus.2012.12.003}.

\bibitem[{Navarro} et~al.(2014){Navarro}, {Madeleine}, {Forget}, {Spiga},
  {Millour}, {Montmessin}, and {M{\"a}{\"a}tt{\"a}nen}]{Nava:14jgr}
T.~{Navarro}, J.-B. {Madeleine}, F.~{Forget}, A.~{Spiga}, E.~{Millour},
  F.~{Montmessin}, and A.~{M{\"a}{\"a}tt{\"a}nen}.
\newblock {Global Climate Modeling of the Martian water cycle with improved
  microphysics and radiatively active water ice clouds}.
\newblock \emph{Journal of Geophysical Research (Planets)}, 2014.
\newblock \doi{10.1002/2013JE004550}.

\bibitem[Ryan and Henry(1979)]{Ryan:79}
J.~A. Ryan and R.~M. Henry.
\newblock Mars atmospheric phenomena during major dust storms as measured at
  surface.
\newblock \emph{J.~Geophys.~Res.}, 84:\penalty0 2821--2829, 1979.

\bibitem[{Spiga} et~al.(2013){Spiga}, {Faure}, {Madeleine},
  {M{\"a}{\"a}tt{\"a}nen}, and {Forget}]{Spig:13rocket}
A.~{Spiga}, J.~{Faure}, J.-B. {Madeleine}, A.~{M{\"a}{\"a}tt{\"a}nen}, and
  F.~{Forget}.
\newblock {Rocket dust storms and detached dust layers in the Martian
  atmosphere}.
\newblock \emph{Journal of Geophysical Research (Planets)}, 118:\penalty0
  746--767, April 2013.
\newblock \doi{10.1002/jgre.20046}.

\bibitem[{Spiga} et~al.(2018){Spiga}, {Banfield}, {Teanby}, {Forget}, {Lucas},
  {Kenda}, {Rodriguez Manfredi}, {Widmer-Schnidrig}, {Murdoch}, {Lemmon},
  {Garcia}, {Martire}, {Karatekin}, {Le Maistre}, {Van Hove}, {Dehant},
  {Lognonné}, {Mueller}, {Lorenz}, {Mimoun}, {Rodriguez}, {Beucler}, {Daubar},
  {Golombek}, {Bertrand}, {Nishikawa}, {Millour}, {Rolland}, {Brissaud},
  {Kawamura}, {Mocquet}, {Martin}, {Clinton}, {Stutzmann}, {Spohn}, {Smrekar},
  and {Banerdt}]{Spig:18insight}
A.~{Spiga}, D.~{Banfield}, N.~A. {Teanby}, F.~{Forget}, A.~{Lucas}, B.~{Kenda},
  J.~A. {Rodriguez Manfredi}, R.~{Widmer-Schnidrig}, N.~{Murdoch}, M.~T.
  {Lemmon}, R.~F. {Garcia}, L.~{Martire}, {\"O}.~{Karatekin}, S.~{Le Maistre},
  B.~{Van Hove}, V.~{Dehant}, P.~{Lognonné}, N.~{Mueller}, R.~{Lorenz},
  D.~{Mimoun}, S.~{Rodriguez}, {\'E}.~{Beucler}, I.~{Daubar}, M.~P. {Golombek},
  T.~{Bertrand}, Y.~{Nishikawa}, E.~{Millour}, L.~{Rolland}, Q.~{Brissaud},
  T.~{Kawamura}, A.~{Mocquet}, R.~{Martin}, J.~{Clinton}, {\'E}.~{Stutzmann},
  T.~{Spohn}, S.~{Smrekar}, and W.~B. {Banerdt}.
\newblock {Atmospheric Science with InSight}.
\newblock \emph{Space Science Reviews}, 214:\penalty0 109, 2018.
\newblock \doi{10.1007/s11214-018-0543-0}.

\bibitem[{Streeter} et~al.(2019){Streeter}, {Lewis}, {Patell}, {Holmes}, and
  {Kass}]{Stre:19}
P.~M. {Streeter}, S.~R. {Lewis}, M.~R. {Patell}, J.~A. {Holmes}, and D.~M.
  {Kass}.
\newblock {Surface Warming During the 2018/MY 34 Mars Global Dust Storm}.
\newblock In \emph{Ninth International Conference on Mars
  2019(LPIContrib.No.2089)}, page 6242, 2019.

\bibitem[Sánchez-Lavega et~al.(2019)Sánchez-Lavega, del Río-Gaztelurrutia,
  Hernández-Bernal, and Delcroix]{Sanc:19storm}
A.~Sánchez-Lavega, T.~del Río-Gaztelurrutia, J.~Hernández-Bernal, and
  M.~Delcroix.
\newblock The onset and growth of the 2018 martian global dust storm.
\newblock \emph{Geophysical Research Letters}, 46\penalty0 (11):\penalty0
  6101--6108, 2019.
\newblock \doi{10.1029/2019GL083207}.
\newblock URL
  \url{https://agupubs.onlinelibrary.wiley.com/doi/abs/10.1029/2019GL083207}.

\bibitem[{Vincendon} et~al.(2015){Vincendon}, {Audouard}, {Altieri}, and
  {Ody}]{Vinc:15}
M.~{Vincendon}, J.~{Audouard}, F.~{Altieri}, and A.~{Ody}.
\newblock {Mars Express measurements of surface albedo changes over 2004-2010}.
\newblock \emph{Icarus}, 251:\penalty0 145--163, 2015.
\newblock \doi{10.1016/j.icarus.2014.10.029}.

\bibitem[{Wang} et~al.(2018){Wang}, {Forget}, {Bertrand}, {Spiga}, {Millour},
  and {Navarro}]{Wang:18}
C.~{Wang}, F.~{Forget}, T.~{Bertrand}, A.~{Spiga}, E.~{Millour}, and
  T.~{Navarro}.
\newblock {Parameterization of Rocket Dust Storms on Mars in the LMD Martian
  GCM: Modeling Details and Validation}.
\newblock \emph{Journal of Geophysical Research (Planets)}, 123:\penalty0
  982--1000, 2018.
\newblock \doi{10.1002/2017JE005255}.

\bibitem[Wilson and Hamilton(1996)]{Wils:96}
R.~W. Wilson and K.~Hamilton.
\newblock Comprehensive model simulation of thermal tides in the {Martian}
  atmosphere.
\newblock \emph{J.~Atmos.~Sci.}, 53:\penalty0 1290--1326, 1996.

\bibitem[{Withers}(2012)]{With:2012}
Paul {Withers}.
\newblock {Empirical Estimates of Martian Surface Pressure in Support of the
  Landing of Mars Science Laboratory}.
\newblock \emph{Space Science Reviews}, 170\penalty0 (1-4):\penalty0 837--860,
  Sep 2012.
\newblock \doi{10.1007/s11214-012-9876-2}.

\bibitem[Zurek(1982)]{Zure:82}
R.~W. Zurek.
\newblock Martian great dust storm, an update.
\newblock \emph{Icarus}, 50:\penalty0 288--310, 1982.

\bibitem[Zurek and Martin(1993)]{Zure:93}
Richard~W. Zurek and Leonard~J. Martin.
\newblock Interannual variability of planet-encircling dust storms on {Mars}.
\newblock \emph{J.~Geophys.~Res.}, 98\penalty0 (E2):\penalty0 3247--3259, 1993.

\end{thebibliography}
\end{document}